\documentclass[reprint,aps,prb]{revtex4-2}

\usepackage{chemformula} % Formula subscripts using \ch{}
\usepackage[T1]{fontenc} % Use modern font encodings
\usepackage{lmodern}
\usepackage{braket}
\usepackage{amsmath}
\usepackage{amssymb}
\usepackage{appendix}
\usepackage{hyperref}
\usepackage{epstopdf}
\usepackage{bm}
\usepackage{dcolumn}% Align table columns on decimal point
\usepackage[version=4]{mhchem}

\usepackage{graphicx}
\usepackage[percent]{overpic} 
\usepackage{color}
\usepackage{transparent} 
\usepackage[caption=false]{subfig}
\begin{document} 
	
	\title{The dynamics of skyrmion shrinking}
	
	\author{Frederik Austrup$^1$}
	\author{Wolfgang H{\"a}usler$^2$}
	\author{Michael Lau$^1$}
	\author{Michael Thorwart$^1$}
	\affiliation{$^1$I. Institut f{\"u}r Theoretische Physik, Universit{\"a}t Hamburg, Notkestra\ss{}e 9, 22607 Hamburg, Germany \\
		$^2$ Institut f{\"u}r Physik, Universit{\"a}t Augsburg, Universit{\"a}tsstra\ss{}e 1, 86135 Augsburg, Germany}
	
	\begin{abstract}
When magnetic skyrmions decay, their size in real space decreases in a finite time before they eventually collapse. We construct an effective continuum model and use its dynamics to describe the shrinking behavior of skyrmions before they collapse. Using the Landau-Lifshitz-Gilbert equation and the time derivative of the vector field, we find a set of coupled nonlinear ordinary differential equation for the time dependent effective skyrmion radius and its helicity. In particular, we use a triangular-shaped skyrmion profile of its polar angle. Contrary to the commonly expected simple exponential decrease in size, we reveal a more complicated time dependence, in which the time-dependent radius crosses over from an exponential decay towards a square root decrease, $\sim (t - t_c)^{1/2}$, near a critical time $t_c$ at which it collapses. This critical time is found to depend logarithmically on the lattice constant. In addition, we examine the interplay between the shrinking dynamics and an accompanying transformation through different skyrmion configurations, depending on the various system parameters. The findings are verified by numerical studies on the lattice, supporting the predictions from the theoretical continuum model.
	\end{abstract}
	
	\maketitle
	\section{Introduction}
	\label{sec:Intro}
The emergence of topologically protected solitons in magnetic systems in the form of skyrmions has garnered significant attention in the field of condensed matter physics. These non-collinear magnetic textures arise as solution of nonlinear field equations, first found in the context of nuclear physics \cite{TSkyrme}. They have become a focal point of research in ferromagnetic materials. Skyrmions arise is solid-state materials with strong spin-orbit coupling and broken inversion symmetry, giving rise to the Dzyaloshinskii-Moriya interaction (DMI) that is the stabilizing factor for skyrmions \cite{Moriya}. 

The investigation of magnetic skyrmion dynamics in ferromagnetic materials has revealed intriguing behavior with potential usage as information carriers in technological applications \cite{Tomasello}. Due to their small size, skyrmions can be used to encode binary information, acting as bits \cite{Tomasello, Nagaosa, Leonov, Zhang, Finocchio, Jiang}, which underscores the importance of understanding how they are written and erased. Thus, particular interest in their creation, stability, annihilation  \cite{RommingWiesendangerWritingDeleting,Stier}, and manipulation \cite{Nagaosa,Iwasaki} has developed.

Several studies have examined the stability of skyrmions using the geodesic nudged elastic band (GNEB) method \cite{BesserabGNEB, Rybakov, Lobanov, HeilRoschMasell}. Here, the energetically lowest intermediate state between the skyrmion state and, e.g., the ferromagnetic state is used to identify the energy barrier separating the two states and thus elucidates the creation and annihilation pathways of skyrmions. Furthermore, the collapse of a quantum skyrmion, existing of only quantum mechanical spins, through quantum tunnelling has been investigated  \cite{Vlasov, Derras-ChoukChudnovskyGaraninQuantum}.

The size of an isotropic skyrmion in the simplest form  with a winding number of 1 may be defined by the radial distance from its center at which the out-of-plane magnetization vanishes. It is common to extract it from the skyrmion's radial profile using a 180° domain wall approximation \cite{Braun, KubetzkaWiesendanger, RommingWisendangerDomainWall, Wang}. An unstable skyrmion is generally expected to shrink over time in its radius until the latter reaches the magnitude of the lattice constant. Then, it eventually collapses on a very short time taken to be instantaneous \cite{RohartMiltatThiaville}. The collapse of skyrmions has been studied in relation to their initial size \cite{CaiChudnovskyGaranin} and with respect to the absence or presence of a magnetic field \cite{Derras-ChoukChudnovskyGaranin}. For sufficiently large skyrmions and in the presence of a magnetic field their size is found to shrink exponentially over time, whereas for smaller skyrmions, the collapse time is proportional to the fourth power of their size, with a logarithmic correction \cite{CaiChudnovskyGaranin,Derras-ChoukChudnovskyGaranin}.

Our study focuses on the dynamics of the shrinking of a skyrmion when unstable, before it eventually collapses. For this, we construct an analytical model and accompany the results by numerical micromagnetic simulations. We consider the configuration when a Bloch-type skyrmion is stabilized by the bulk DMI \cite{MuehlbauerBoeni}. We derive a set of coupled ordinary differential equations for the time evolution of the skyrmion radius and helicity, constructed on the basis of  the Landau-Lifshitz-Gilbert (LLG) equation \cite{LandauLifshits,Gilbert} for the vector field of the normalized magnetization, parametrized via a skyrmion profile of triangular shape \cite{BogdanovYablonskii}. For zero DMI, i.e., when the skyrmion is unstable, the differential equation for the skyrmion radius decouples from the equation for the helicity. Solving the differential equation for the radius at vanishing DMI analytically, we find a shrinking behavior of the skyrmion that comprises an initial exponential decay at large radii and a square-root type decay for small radii close to the final collapse at a critical time. This critical time is found to increase logarithmically with decreasing lattice constant. These findings align with numerical studies, in which we systematically extend the accessible parameter range by tuning external parameters, including the applied magnetic field, the Dzyaloshinskii-Moriya interaction, and the Gilbert damping, across various grid densities. The numerical solutions of the time evolution of the skyrmion radius can be divided into three phases. Initially, at very short times, the skyrmion decays quadratically in time, followed by an exponential decay. As the skyrmion radius approaches the lattice constant, a square-root type decay dynamics emerges, culminating in a rapid collapse upon reaching a radius of the order of the lattice constant.

At finite DMI the nonlinear differential equations for radius and helicity are not decoupled and need to be solved numerically. Oscillations are found in the shrinking dynamics of the radius, accompanied by helical rotations of the in-plane magnetization. Numerical simulation results confirm this behavior and reveal further breathing modes during the skyrmion shrinking. Such a general phenomenon has also been analysed \cite{McKeeverEverschor-Sitte} in a Hamiltonian model to describe the dynamics of skyrmions and antiskyrmions using collective coordinates. For skyrmion breathing modes, characterized by periodic expansion and contraction of the skyrmion radius, two main types of dynamics are known: a small oscillation mode and a rotation mode in which the magnetization continuously switches between N\'eel- and Bloch-type configurations. For higher energies, the breathing dynamics enters a rotational phase, with significant differences in energy dissipation rates. Our work complements this study by focusing on the shrinking dynamics up to the eventual collapse of an unstable skyrmion in a system with bulk DMI.
	
\section{Model}
\label{sec:Model}
    We study the shrinking of a skyrmion in a ferromagnet where skyrmions are not the energetically favorable state. The ensuing dynamics of the normalized magnetic moments $\bm{n}_i=\bm{n}_i(x,y,t)$ on a two-dimensional lattice at lattice sites $i$ are described by the LLG equation \cite{LandauLifshits,Gilbert} as
\begin{equation}
\label{eq:LLGlattice}
\frac{\partial \bm{n}_i}{\partial t}=-\bm{n}_i \times \bm{B}_i^{\text{eff}} + \alpha \bm{n}_i \times \frac{\partial \bm{n}_i}{\partial t},
\end{equation}
where the first term on the right hand side describes the precessional motion of $\bm{n}$ around a local effective magnetic field $\bm{B}_i^{\text{eff}}$ and the second term describes the energy dissipation that damps the motion to eventually align with the effective field. The damping is parametrized via the phenomenological Gilbert damping parameter $\alpha$. We adjust the gyrocoupling constant $\gamma$ in the following in such way that it ensures a time unit as $t_0 = 1/J$.

The system parameters are the exchange constant $J$, the DMI-strength $D$, and the external magnetic field $B$ pointing in $z$-direction. They are introduced in the lattice-Hamiltonian
\begin{equation}
\label{eq:Hamiltonian_general}
H = -\frac{J}{2} \sum_{\braket{i,j}} \bm{n}_i \bm{n}_j - D \sum_{\braket{i,j}} \bm{d}_{ij} \left[ \bm{n}_i \times \bm{n}_j \right] - B \sum_i n_i^z,
\end{equation}
that defines the effective magnetic field at each lattice site as $\bm{B}_i^{\text{eff}}=-\partial H/\partial \bm{n}_i$. Summation $\braket{i,j}$ includes nearest neighbors only. The DMI of strength $D$ here is chosen to favor Bloch-type skyrmions, with the tensor $\bm{d}_{ij} = \frac{\bm{r}_j-\bm{r}_i}{|\bm{r}_j-\bm{r}_i|}$ with lattice positions $\bm{r}_{j}=(x_j, y_j)$ in the two-dimensional space \cite{Leonov}.

\section{Continuum Theory}

In the following we study the shrinking dynamics of a skyrmion analytically within a continuum theory \cite{KronmuellerFaehnle}. The Hamiltonian of Eq.\ \eqref{eq:Hamiltonian_general} and the LLG equation in Eq.\ \eqref{eq:LLGlattice} are expressed in terms of a classical field theory for the parametrized vector field of a skyrmion. Furthermore, the corresponding density of the topological charge is considered.

\subsection{Energy functional of a skyrmion}
\label{sec:blochtypesk}

To replace the magnetic moments in Eq.\ \eqref{eq:Hamiltonian_general} by a continuous vector field a Taylor expansion of the position vectors is used in space. Assuming a square lattice, the expansion is performed around the midpoint $k$ between the lattice sites $i$ and $j$. The resulting vectors of the magnetic moments are of the form $\bm{n}_{\bm{r}_k \pm \frac{a}{2}\textbf{e}_s}$, with unit vectors $\textbf{e}_s$ and spatial indices $s=\{ x,y \}$. Applying the Taylor expansion, the magnetic moments read $\bm{n}_{\bm{r}_k \pm \frac{a}{2}\textbf{e}_s} \approx \bm{n}_{\bm{r}_k} \pm \frac{a}{2} \partial_s \bm{n}_{\textbf{r}_k}$. The exchange contribution of the Hamiltonian then reads
\begin{equation}
-\frac{J}{2} \sum_{\braket{i,j}} \bm{n}_i \bm{n}_j \approx \text{const} + \frac{J a^2}{2} \sum_{k,s} \left( \partial_s \bm{n} \right)^2,
\end{equation}
where the indices for all of the $\bm{n}_{\bm{r}_k}$ have been dropped for readability. Also, the constant will be neglected from here on. The Dzyaloshinskii-Moriya contribution can be treated in the same way and we obtain 
\begin{equation}
 - D \sum_{\braket{i,j}} \bm{d}_{ij} \left[ \bm{n}_i \times \bm{n}_j \right] = -aD \sum_{k,s} \bm{n} \times \partial_s \bm{n}.
\end{equation}

In the continuum limit we replace $\sum_k \longrightarrow \int \text{d}^2r/a^2$, yielding for each term of the Hamiltonian

\begin{equation}
\label{eq:LatticeToContinuum}
\begin{split}
\frac{Ja^2}{2} \sum_{k,s} \left( \partial_s \bm{n} \right)^2 &~ \longrightarrow \frac{J}{2} \int \text{d}^2r \left[ \left( \partial_x \bm{n} \right)^2 + \left( \partial_y \bm{n} \right)^2 \right], \\
-aD \sum_{k,s} \bm{n} \times \partial_s \bm{n} &~ \longrightarrow -\frac{D}{a} \int \text{d}^2r ~ \bm{n} \left( \nabla \times \bm{n} \right), \\
-B\sum_i n_i^z &~ \longrightarrow -\frac{B}{a^2} \int  \text{d}^2r ~n_z.
\end{split}
\end{equation}
At this point, $D/a$ and $B/a^2$ are redefined as $\tilde{D}$ and $\tilde{B}$, respectively, thereby absorbing the lattice constant. The energy functional then reads
\begin{equation}
\label{eq:EnergyFunctional}
\begin{split}
E=\int \text{d}^2r & \left[ \frac{J}{2} \left[ \left( \partial_x \bm{n} \right)^2 + \left( \partial_y \bm{n} \right)^2 \right] \right. \\
& \left. \vphantom{\frac{J}{2}} - \tilde{D} \bm{n} \left( \nabla \times \bm{n} \right) - \tilde{B} n_z \right].
\end{split}
\end{equation}
From this, the effective magnetic field $\bm{B}_\text{eff}$ is given in the continuum by the functional derivative of the energy density $\mathcal{E}$ with respect to $\bm{n}$ as
\begin{equation}
\label{eq:Beff}
\bm{B}_{\text{eff}} = -\frac{\delta \mathcal{E}}{\delta \bm{n}} = J \Delta \bm{n} - 2 \tilde{D} \left(\nabla \times \bm{n} \right) + \tilde{B} \left( \begin{array}{c}
0 \\ 
0 \\ 
1
\end{array} \right),
\end{equation}
where $\mathcal{E}$ is the integrand of the energy functional Eq.\ \eqref{eq:EnergyFunctional}. Now, the LLG equation in the continuum limit reads
\begin{equation}
\label{eq:LLG}
\frac{\partial}{\partial t} \bm{n} = - \bm{n} \times \bm{B}_\text{eff}a^2 + \alpha \bm{n} \times \frac{\partial}{\partial t} \bm{n},
\end{equation}
in agreement with Ref.\ \cite{Schuette}, where the authors derive the continuum LLG equation from the dynamic part of a Lagrangian via the Berry phase, which includes the lattice constant.

Expressed by spherical coordinates, the magnetisation vector is given by $\bm{n}=\left( \sin\Theta\cos\Phi, \sin\Theta\sin\Phi, \cos\Theta \right)^\texttt{T}$. Taking the center of the skyrmion as origin of the coordinate system, the polar angle $\Theta(\rho)$ is a function of the distance $\rho$ from the center, while the azimuthal angle is given as $\Phi=m\varphi+\varphi_0$ with $\varphi$ being the azimuthal coordinate and the helicity $\varphi_0$ that governs the skyrmions configuration. A Néel-type skyrmion is formed with $m=1$ and $\varphi_0=0$ and a Bloch-type skyrmion with $m=1$ and $\varphi_0=\pi/2$. In all our studies, the skyrmion is initially prepared in the Bloch-type configuration that is favoured by the given bulk DMI. The energy density $\mathcal{E} =\mathcal{E} (\rho,\Theta(\rho),\Phi(\varphi))$ for an arbitrary skyrmion configuration reads
\begin{equation}
\label{eq:EnergyDensity}
\begin{split}
\mathcal{E} =& \frac{J}{2} \left[ \Theta'^2+\frac{\sin^2\Theta}{\rho^2} \right] + \tilde{D} \left[ \Theta'+\frac{\sin(2\Theta)}{2\rho} \right]\sin\varphi_0 \\
& - \tilde{B} \cos\Theta \, ,
\end{split}
\end{equation}
where the prime indicates the derivative with respect to $\rho$, i.e., $\Theta'=\text{d}\Theta/\text{d}\rho$.

\subsection{Topological charge of a skyrmion}
\label{sec:WindingNumber}
The winding number, also known as the topological charge or skyrmion number, counts how many times the field configuration wraps around the unit sphere and is defined as the integral over the topological charge density \cite{Everschor-Sitte}
\begin{equation}
\label{eq:chargeDensity}
q=\bm{n} \cdot \left( \frac{\partial \bm{n}}{\partial x} \times \frac{\partial \bm{n}}{\partial y} \right)
\end{equation}
as
\begin{equation}
\label{eq:topCharge}
\mathcal{Q}=\frac{1}{4 \pi} \int \int \bm{n} \cdot \left( \frac{\partial \bm{n}}{\partial x} \times \frac{\partial \bm{n}}{\partial y} \right)  \text{d}x \text{d}y, ~~~\mathcal{Q} \in \mathbb{Z}.
\end{equation}
Skyrmions of topological charges up to $\mathcal{Q}=10$ have been observed experimentally \cite{Albrecht24}. The charge density $q$ is independent of the azimuthal angle and reads for the parametrisation describing a $|\mathcal{Q}|=1$ skyrmion
\begin{equation}
\label{eq:blochCharge}
q=\frac{\Theta'}{\rho}\sin\Theta.
\end{equation}

\section{Analytical treatment for a triangular skyrmion}
Fig. \ref{fig:Triangular-Ansatz} shows the radial profile $\Theta(\rho)$ of the polar angle of a Bloch-type skyrmion, starting at the center of the skyrmion (spin down) with a smooth transition into the ferromagnetic background (spin up). The exact functional dependence $\Theta(\rho)$ cannot be expressed by known functions \cite{BogdanovHubert}. Often explicit functions are used, fitting the exact dependence quite well \cite{Wang}. One frequently used form is
\begin{equation}
\label{eq:RadiusFit}
\Theta(\rho)=2\arctan\left[ \frac{\sinh(\rho_0 / w)}{\sinh(\rho / w)} \right],
\end{equation}
where $\rho_0$ is the skyrmion radius and $w$ is the width of the domain wall-like shoulder seen in typical profiles.

In the following, we use an even more simplistic triangular form of the profile which has been proposed by Bogdanov and Yablonskii \cite{BogdanovYablonskii}, due to the fact that it allows for analytic treatment. This approximation is shown as the red line in Fig.\ \ref{fig:Triangular-Ansatz} (a) and enables us to find explicit expressions for the energy functional and renders the LLG equation tractable. By this, we are able to derive a set of two coupled  nonlinear ordinary differential equations for the time evolution of the radius and the helicity of the triangular shaped skyrmion.

\subsection{Triangular approximation of the polar angle}
\begin{figure}[t!]
	\centering
	\includegraphics[width=0.47\textwidth]{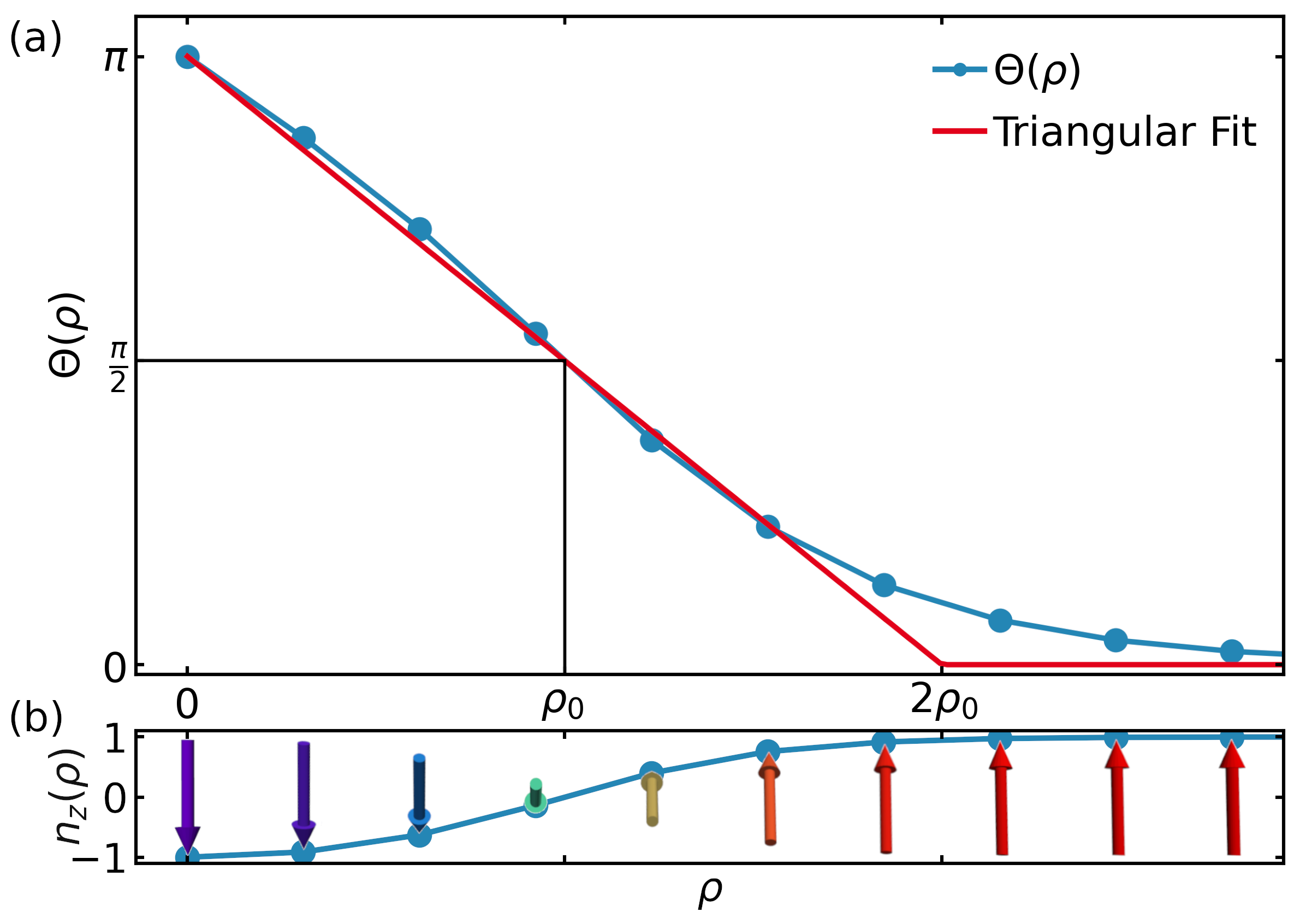}
	\caption{Visualisation of the radial profile of a Bloch-type skyrmion. 
	(a) $\Theta(\rho)$-profile (blue solid line with dots), 	(b) $z$-component of  $\bm{n}$ (blue line with symbols) as a function of the radial distance $\rho$ (given in units of the lattice constant) from the skyrmion center placed at $\rho=0$. The arrows indicate the projected magnetic moments. The ferromagnetic background  appears for $\rho \to \infty$.  The skyrmion radius is defined by the radial position for which the $z$-component of the vector field vanishes, i.e., $n_z(\rho_0)=0$ such that $\Theta(\rho)=\pi /2$, see black solid rectangle in (a). The red line in (a) depicts the profile of a triangular skyrmion of Eq.\ \eqref{eq:heaviside} approximating  the real profile. 
		\label{fig:Triangular-Ansatz}}
\end{figure}

Using the Heaviside function $\mathcal{H}(x)$ the triangular form of the radial profile is given by
\begin{equation}
\label{eq:heaviside}
\Theta(\rho)=(\pi - \xi)\mathcal{H}(\pi - \xi), ~~~\text{with}~\xi = \frac{\pi \rho}{2\rho_0}.
\end{equation}
Here, $\rho_0$ is the skyrmion radius, defined by the relation $n_z(\rho_0)=0$, i.e., by that $\rho=\rho_0$ where the magnetization vector lies in  the $x$-$y$-plane as shown in Fig.\ \ref{fig:Triangular-Ansatz}. Using Eq.\ \eqref{eq:heaviside}, the magnetic vector field then is parametrised as
\begin{equation}
\label{eq:nvecBlochTri}
\bm{n}= \left( \begin{array}{c}
\sin\xi\cos\Phi \\ 
\sin\xi\sin\Phi \\ 
-\cos\xi
\end{array} \right).
\end{equation}

\subsection{Energy functional of a triangular Bloch skyrmion}

The energy density in Eq.\  \eqref{eq:EnergyDensity} can be expressed in terms of the triangular skyrmion parametrization as

\begin{equation}
\label{eq:TriangleEnergyDensity}
\begin{split}
\mathcal{E}&(\xi, \rho_0, \varphi_0)= \frac{J}{2} \left[ \frac{\pi^2}{4\rho_0^2} + \frac{\pi^2}{4\rho_0^2} \frac{\sin^2\xi}{\xi^2} \right] \\
& - \tilde{D} \left[ \frac{\pi}{2\rho_0} + \frac{\pi}{2\rho_0} \frac{\sin(2\xi)}{2\xi} \right]\sin\varphi_0 + \tilde{B} \cos\xi + \tilde{B}.
\end{split}
\end{equation}
The last term is added to subtract the ferromagnetic (FM) ground state energy density from the energy density determined above. 
The total energy follows as 
\begin{equation}
E=E(\rho_0,\varphi_0)= \frac{4\rho_0^2}{\pi^2} \int_0^{2\pi} \int_0^\pi \xi \, \mathcal{E} (\xi, \rho_0, \varphi_0) ~ \text{d}\xi \text{d}\varphi\, ,
\end{equation}
yielding the energy being only a function of the skyrmion radius $\rho_0$ and its helicity $\varphi_0$ in the explicit form
\begin{equation}
\label{eq:TriangularEnergy}
\begin{split}
E(\rho_0,\varphi_0)=& J \left[ \frac{\pi^3}{2} + \frac{\pi}{2} \left( \ln(2\pi) + C - \texttt{Ci}(2\pi)\right) \right] \\
& - 2 \pi^2 \rho_0 \tilde{D} \sin\varphi_0 + \left(\underbrace{\pi}_{\text{FM}} - \frac{4}{\pi} \right) 4 \rho_0^2 \tilde{B},
\end{split}
\end{equation}
with the Euler-Mascheroni constant $C$ and the cosine integral $\texttt{Ci}$ \cite{Gradshteyn}. The expression in the square bracket can be evaluated to $19.3322$.

\begin{figure}[t!]
	\centering
	\includegraphics[width=0.47\textwidth]{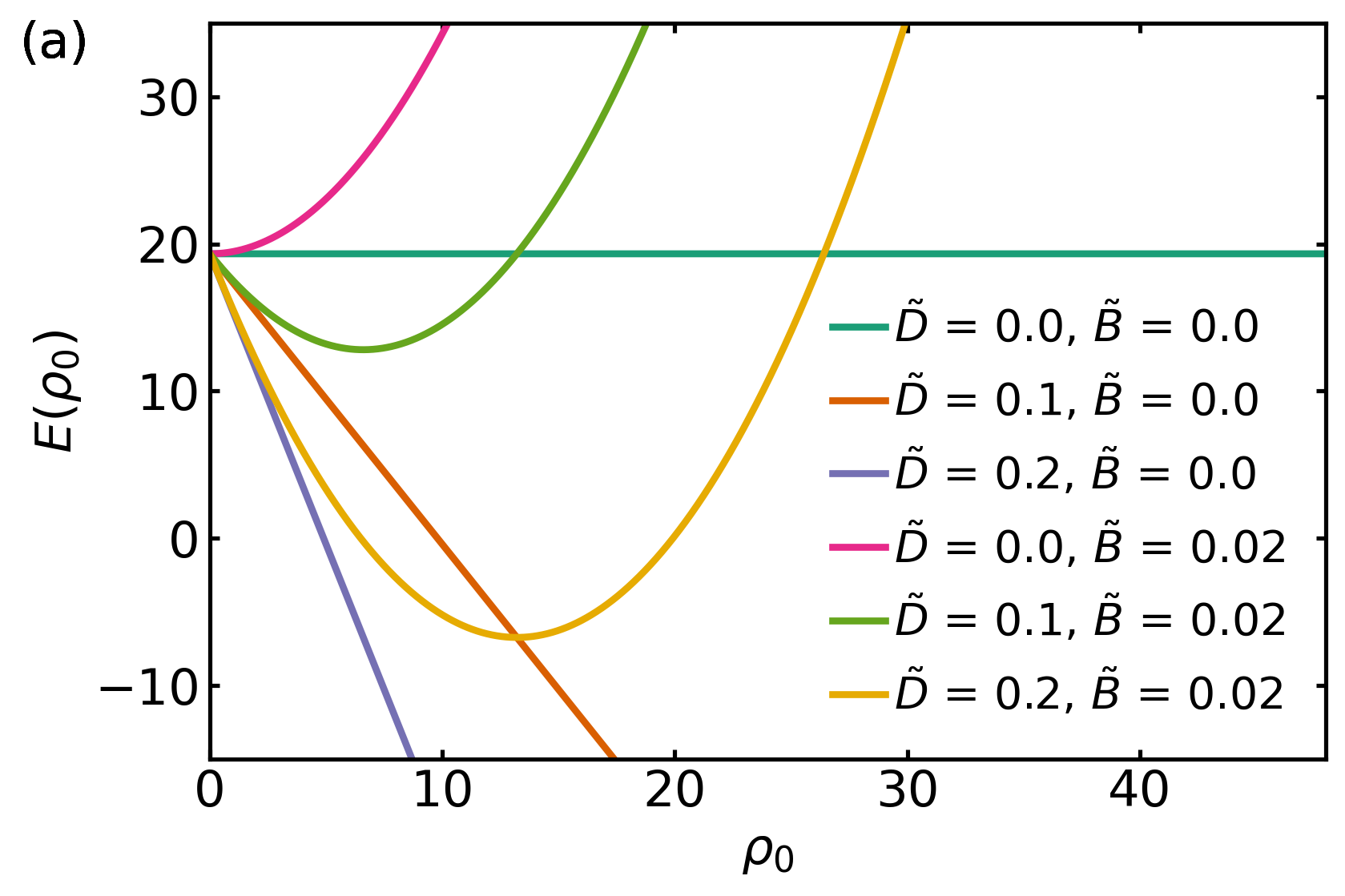}
	\includegraphics[width=0.47\textwidth]{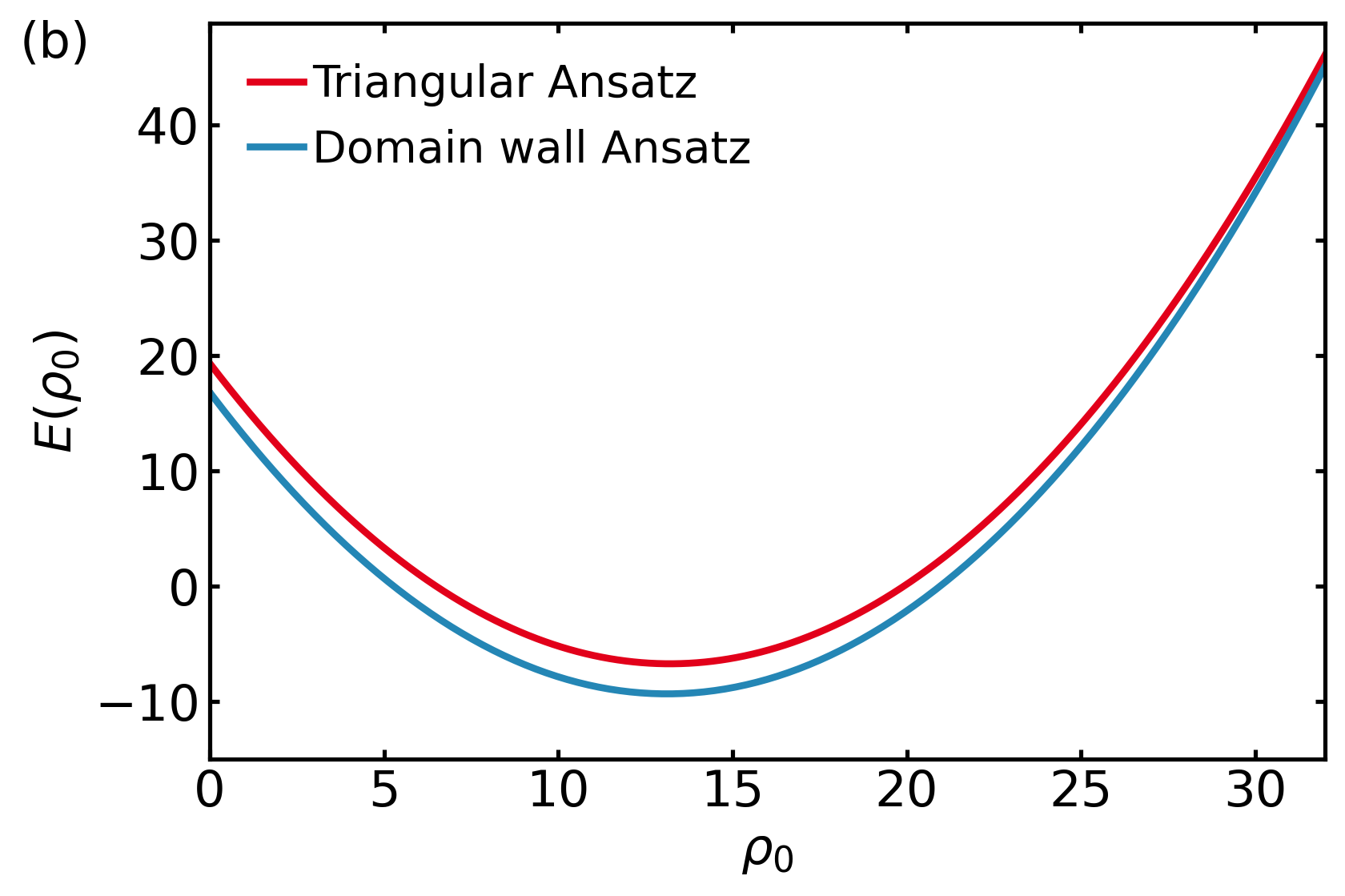}
	\caption{(a) Total energy of the triangular Bloch-type skyrmion derived from the energy density in Eq.\ (\ref{eq:TriangleEnergyDensity}) with $\varphi_0=\pi/2$ in dependence of the skyrmion radius $\rho_0$ for different values of $\tilde{D}$ and $\tilde{B}$ (all given in units of $J$). (b) Comparison of the total energy using the triangular skyrmion profile and the total energy calculated numerically using the domain wall skyrmion in Eq.\ \eqref{eq:RadiusFit} with $\tilde{D}=0.2$ and $\tilde{B}=0.02$.
		\label{fig:TotalEnergy}}
\end{figure}

Fig.\ \ref{fig:TotalEnergy} (a) shows the dependence of $E$ on the radius $\rho_0$ of a Bloch-type skyrmion governed by $\varphi_0=\pi/2$ for different values of $\tilde{B}$ and $\tilde{D}$, all given in units of $J$. Notably, the exchange contribution stays independent of $\rho_0$. The contribution of the DMI diverges to negative energies. With an external magnetic field an energy minimum forms, stabilizing the skyrmion. The skyrmion radius $\rho_0^{\text{st}}$, for which the energy of the triangular skyrmion is minimal, is then given by
\begin{equation}
\label{eq:energyMinimum}
\rho_0^{\text{st}}=\frac{\pi^3\tilde{D}}{4\left( \pi^2 - 4 \right) \tilde{B}}.
\end{equation}

This equation coincides with that for the energy minimizing radius in Ref. \cite{NavauSanchez} for, in our case, vanishing anisotropy. To understand the accuracy of the triangular approximation regarding the total energy, we compare its dependence on the skyrmion radius to that of the energy that the domain wall approximation of Eq.\ \eqref{eq:RadiusFit} provides. The comparison is depicted in Fig.\ \ref{fig:TotalEnergy} (b) where the latter was obtained by numerical integration of Eq.\ \eqref{eq:EnergyDensity} with Eq.\ \eqref{eq:RadiusFit}. In general, the triangular Ansatz (red line) slightly overestimates the energy of the more accurate domain wall approach, but the general form and, especially, the energetically most favorable skyrmion radius are recovered quite well by the triangular Ansatz. Thus, we conclude that the qualitative agreement of the energies of a triangular and a domain wall skyrmion suggests that the triangular approach should be a trusty approximation for deriving the dynamics.

\begin{figure}[t!]
	\centering
	\subfloat{\includegraphics[width=0.47\textwidth]{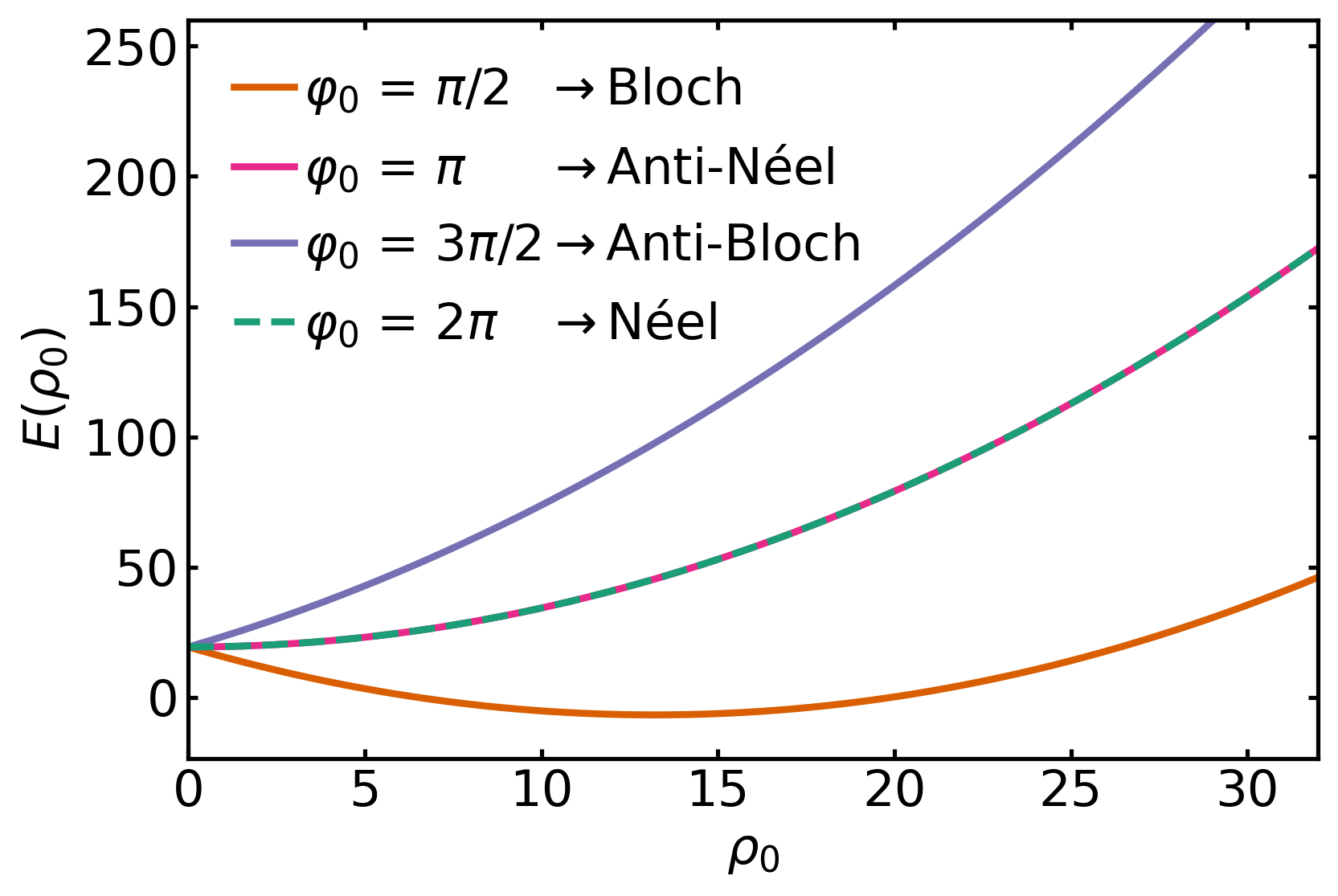}}
	\caption{Comparison of the total energy of the different skyrmion configurations in a system with bulk DMI at $\tilde{D}=0.2$ and $\tilde{B}=0.02$. This favors the Bloch-type skyrmion, while the anti-N\'eel- and N\'eel-type configurations are degenerate and higher in energy and the anti-Bloch-type skyrmion has the highest energy.
		\label{fig:EnergyPhi0}}
\end{figure}

Furthermore, Fig.\ \ref{fig:EnergyPhi0} shows the dependence of the total energy for different helicities of the skyrmion. Given the bulk DMI in Eq.\ \eqref{eq:TriangularEnergy}, the Bloch-type skyrmion is the configuration yielding an energy minimum at a finite skyrmion radius ($\sin(\pi/2)=1$). The anti-N\'eel- and N\'eel-type skyrmions are degenerate and higher in energy ($\sin(\pi)=\sin(2\pi)=0$), while the anti-Bloch-type configuration yields the highest energy ($\sin(3\pi/2)=-1$).

%\subsection{Topological charge density}
%The charge density of a skyrmion in Eq.\ \eqref{eq:blochCharge} can be evaluated in the case of the triangular skyrmion Ansatz in Eq.\ \eqref{eq:heaviside} to the form
%%
%\begin{equation}
%q=-\frac{\pi^2}{4\rho_0^2}\frac{\sin\xi}{\xi}=-\frac{\pi}{2\rho \rho_0} \sin\xi, \quad \text{for}~\rho<2\rho_0 \, .
%\end{equation}

\subsection{Differential equation for the skyrmion radius}
With the triangular Ansatz we parametrize the skyrmion by its radius $\rho_0$ and its helicity $\varphi_0$. In general, we study a spatially stationary but unstable skyrmion. Therefore, these quantities become time dependent, $\rho_0(t)$ and $\varphi_0(t)$. For an equation of motion, we calculate the derivative of the vector field $\bm{n}$ in Eq.\ \eqref{eq:nvecBlochTri} with respect to time as
\begin{equation}
\label{eq:DotVecField}
\dot{\bm{n}}(\rho_0(t), \varphi_0(t))= \left( \begin{array}{c}
-\xi \frac{\dot{\rho}_0}{\rho_0} \cos\xi  \cos\Phi -  \dot{\varphi}_0 \sin\xi\sin\Phi \\ 
-\xi \frac{\dot{\rho}_0}{\rho_0} \cos\xi  \sin\Phi +  \dot{\varphi}_0 \sin\xi\cos\Phi \\ 
-\xi \frac{\dot{\rho}_0}{\rho_0} \sin\xi
\end{array} \right).
\end{equation}

Furthermore, we use Eq.\ \eqref{eq:nvecBlochTri} to derive an expression for the effective magnetic field in Eq.\ \eqref{eq:Beff}, which allows us to express the LLG in Eq.\ \eqref{eq:LLG} in terms of the triangular skyrmion. These expressions are too long to be displayed here, but they depend on $\rho$, $\rho_0$, $\varphi$ and $\varphi_0$. Since the Skyrmion does not move spatially, we integrate out the spatial coordinate $\varphi$ and find for the $x$- and $y$-components
\begin{equation}
\label{eq:zeros}
\begin{split}
\int_0^{2\pi} d\varphi \, [\dot{\bm{n}}]_x &= 0, \\
\int_0^{2\pi} d\varphi \, [\dot{\bm{n}}]_y &= 0.
\end{split}
\end{equation}
However, integrating the $z$-component gives a contribution reading
\begin{equation}
\label{eq:LLGzdphi}
\begin{split}
\int_0^{2\pi} d\varphi \, [\dot{\bm{n}}]_z = &\frac{a^2\alpha}{1+\alpha^2} \left( 2\pi \tilde{B} \sin^2\xi \vphantom{\frac{\sin^2\xi}{\xi}} \right. \\
+ \tilde{D} \frac{2\pi^2}{\rho_0} & \left[ -\sin^2\xi\frac{1}{\alpha} \cos\varphi_0 - \frac{\sin^3\xi}{\xi}\sin\varphi_0 \right] \\
+ J \frac{\pi^3}{2\rho_0^2} & \left.\left[ \frac{\sin\xi}{\xi} - \frac{\sin^2\xi \cos\xi}{\xi^2} \right] \right).
\end{split}
\end{equation}
We also can multiply the $x$- and $y$-components of the LLG by $\cos\Phi$ and $\sin\Phi$, respectively. Integrating over $\varphi$ then yields
\begin{equation}
\label{eq:LLGxydphi}
\begin{split}
\int_0^{2\pi}& d\varphi \, [\dot{\bm{n}}]_x \cdot \cos\Phi = \int_0^{2\pi} d\varphi \, [\dot{\bm{n}}]_y \cdot \sin\Phi \\
& = \frac{a^2\alpha}{1+\alpha^2} \left( \pi \tilde{B} \sin\xi\cos\xi \vphantom{\frac{\sin^2\xi}{\xi}} \right. \\
+ \tilde{D} \frac{\pi^2}{\rho_0} & \left[ -\sin\xi\cos\xi\frac{1}{\alpha} \cos\varphi_0 - \frac{\sin^2\xi\cos\xi}{\xi}\sin\varphi_0 \right] \\
+ J \frac{\pi^3}{4\rho_0^2} & \left.\left[ \frac{\cos\xi}{\xi} - \frac{\sin\xi \cos^2\xi}{\xi^2} \right] \right).
\end{split}
\end{equation}

With the corresponding steps applied to the components of Eq.\ \eqref{eq:DotVecField}, we get for the $z$-component
\begin{equation}
\label{eq:ndotzdphi}
\int_0^{2\pi} \, d\varphi [\dot{\bm{n}}]_z = -2\pi \frac{\dot{\rho}_0}{\rho_0}\xi\sin\xi,
\end{equation}
and for the $x$- and $y$-components
\begin{equation}
\label{eq:ndotxydphi}
\int_0^{2\pi} d\varphi \, [\dot{\bm{n}}]_x \cdot \cos\Phi = \int_0^{2\pi} d\varphi \, [\dot{\bm{n}}]_y \cdot \sin\Phi  = -\pi \frac{\dot{\rho}_0}{\rho_0}\xi\cos\xi.
\end{equation}

Equating Eq.\ \eqref{eq:LLGzdphi} and Eq.\ \eqref{eq:ndotzdphi}, or, equivalently, Eq.\ \eqref{eq:LLGxydphi} and Eq.\ \eqref{eq:ndotxydphi} yields the nonlinear differential equation
\begin{equation}
\label{eq:rho0ODE}
\begin{split}
\dot{\rho}_0\xi=&\frac{a^2\alpha}{1+\alpha^2} \left( - \rho_0 \tilde{B} \sin\xi \vphantom{\frac{\sin\xi}{\xi}} \right. \\
&  + \tilde{D} \pi \left[ \sin\xi\frac{1}{\alpha} \cos\varphi_0 + \frac{\sin^2\xi}{\xi}\sin\varphi_0 \right] \\
& \left. -J\frac{\pi^2}{4\rho_0} \left[ \frac{1}{\xi} - \frac{\sin\xi \cos\xi}{\xi^2} \right] \right).
\end{split}
\end{equation}

Essential to the desired form of the equations for $\dot{\bm{n}}$ is our assumption that $\Theta(\rho)$ should maintain a triangular shape over time. This requirement dictates that no $\xi$-dependence should appear in Eq.\ \eqref{eq:rho0ODE}, or, in other words, that the $\xi$-dependence on the left hand side must equal the $\xi$-dependence on the right hand side as a force term. Other $\xi$-dependencies would alter the skyrmion shape in time. Therefore, only functional components should be considered on the right hand side that align with the $\xi$-dependence on the left side --- all components orthogonal to this have to be disregarded. Projection upon this $\xi$-dependence yields the scalar products
\begin{equation}
\begin{split}
c_1=&\frac{\int_0^\pi \text{d}\xi \xi^2 \sin\xi}{\int_0^\pi \xi \text{d}\xi \xi^2} = \frac{\pi^2-4}{\pi^4/4} \approx 0.241029, \\
c_2=&\frac{\pi\int_0^\pi \text{d}\xi \xi^2 \sin\xi}{\int_0^\pi \xi \text{d}\xi \xi^2} = \pi c_1 \approx 0.757215, \\
c_3=&\frac{\pi\int_0^\pi \text{d}\xi \xi \sin^2\xi}{\int_0^\pi \xi \text{d}\xi \xi^2} = \frac{\pi^3/4}{\pi^4/4} = \frac{1}{\pi} \approx 0.318310, \\
c_4=&\frac{\frac{\pi^2}{4}\int_0^\pi \text{d}\xi \left( \xi - \sin\xi \cos\xi \right)}{\int_0^\pi \xi \text{d}\xi \xi^2} = \frac{\pi^4/8}{\pi^4/4} = 0.5,
\end{split}
\end{equation}

Finally, a nonlinear ordinary differential equation can be obtained for $\rho_0(t)$
\begin{equation}
\label{eq:finalODErho}
\begin{split}
\dot{\rho}_0=& \frac{a^2\alpha}{1+\alpha^2} \left(- c_1 \tilde{B} \rho_0 + c_3 \tilde{D} \sin\varphi_0 - c_4 J \frac{1}{\rho_0} \right) \\
&+ \frac{a^2}{1+\alpha^2}c_2 \tilde{D} \cos\varphi_0,
\end{split}
\end{equation}
with the right-hand side depending on $\varphi_0(t)$.

\subsection{Differential equation for the skyrmion helicity}
To obtain a differential equation for the skyrmion helicity $\varphi_0(t)$ from the $x$- and $y$-components of Eq.\ \eqref{eq:DotVecField} and the LLG equation, the $x$- and $y$-components are multiplied by $\sin\Phi$ and $\cos\Phi$, respectively, and then integrated over $\varphi$. This leads to
\begin{equation}
\label{eq:dotnztimessinPhialpha0}
\int_0^{2\pi} d\varphi \, [\dot{\bm{n}}]_x \cdot \sin\Phi = - \int_0^{2\pi} d\varphi \, [\dot{\bm{n}}]_y \cdot \cos\Phi = - \dot{\varphi}_0 \pi \sin\xi
\end{equation}
and
\begin{equation}
\begin{split}
\int_0^{2\pi} d\varphi \, &[\dot{\bm{n}}]_x \cdot \sin\Phi = - \int_0^{2\pi} d\varphi \, [\dot{\bm{n}}]_y \cdot \cos\Phi \\
 =& \frac{a^2}{1+\alpha^2} \left( -\tilde{B} \pi \sin\xi \vphantom{\frac{\sin^2\xi}{\xi}} \right. \\
& + \pi^2 \tilde{D} \frac{1}{\rho_0} \left[ -\alpha \sin\xi\cos\varphi_0 + \frac{\sin^2\xi}{\xi}\sin\varphi_0  \right] \\
& \left. + J \frac{\pi^3}{4\rho_0^2} \left[ -\frac{1}{\xi} + \frac{\sin\xi \cos\xi}{\xi^2} \right] \right).
\end{split}
\end{equation}
Equating these two equations gives
\begin{equation}
\label{eq:phi0equation}
\begin{split}
\dot{\varphi}_0 \sin\xi =& \frac{a^2}{1+\alpha^2} \left( \tilde{B} \sin\xi \vphantom{\frac{\sin^2\xi}{\xi}}  \right. \\
&  - \pi \tilde{D} \frac{1}{\rho_0} \left[ -\alpha \sin\xi\cos\varphi_0 + \frac{\sin^2\xi}{\xi}\sin\varphi_0 \right] \\
& \left. + J \frac{\pi^2}{4\rho_0^2} \left[ \frac{1}{\xi} - \frac{\sin\xi \cos\xi}{\xi^2} \right] \right).
\end{split}
\end{equation}
Assuming an isotropic skyrmion helicity at any time suggests that $\varphi_0(t)$ must stay independent of $\rho$. Therefore, we project Eq.\ \eqref{eq:phi0equation} on the $\xi$-dependence of its left hand side, as above. This yields the scalar products
\begin{equation}
\begin{split}
c_5 =& \frac{\int_0^\pi \text{d}\xi \xi \sin^2\xi}{\int_0^\pi \xi \text{d}\xi \sin^2\xi} = 1, \\
c_6 =& \frac{\pi \int_0^\pi \text{d}\xi \xi \sin^2\xi}{\int_0^\pi \xi \text{d}\xi \sin^2\xi} = \pi, \\
c_7 =& \frac{\pi \int_0^\pi \text{d}\xi \sin^3\xi}{\int_0^\pi \xi \text{d}\xi \sin^2\xi} = \frac{16}{3\pi} \approx 1.697653, \\
c_8 =& \frac{\frac{\pi^2}{4}\int_0^\pi \text{d}\xi \left( \sin\xi - \frac{\sin^2\xi \cos\xi}{\xi} \right)}{\int_0^\pi \xi \text{d}\xi \sin^2\xi} \approx 1.709585.
\end{split}
\end{equation}
The ordinary differential equation for $\varphi_0(t)$ thus becomes
\begin{equation}
\label{eq:finalODEphi}
\begin{split}
\dot{\varphi}_0=\frac{a^2}{1+\alpha^2} & \left(c_5 \tilde{B} - \tilde{D} \frac{1}{\rho_0}\left[ -\alpha c_6 \cos\varphi_0 +  c_7 \sin\varphi_0 \right] \right. \\
& \left. + c_8 J \frac{1}{\rho_0^2} \right)  \, ,
\end{split}
\end{equation}
whose right-hand side depends on $\rho_0(t)$.

\subsection{Solution of the coupled differential equations}
For large skyrmions ($c_1 \tilde{B}\rho_0 \gg c_4 J / \rho_0$) the equation of motion \eqref{eq:finalODErho} yields an exponentially decreasing skyrmion radius
\begin{equation}
\rho_0 \sim \rho_{0,0} \exp\left[-\frac{c_1a^2\alpha}{1+\alpha^2}\tilde{B}t \right],
\end{equation}
where $\rho_{0,0}$ is the skyrmion radius at $t=0$. A square root behavior then takes over for small skyrmion sizes ($c_4 J / \rho_0 \gg c_1 \tilde{B} \rho_0$)
\begin{equation}
\label{eq:squareroot}
\rho_0 \sim \sqrt{\frac{2c_4a^2\alpha}{1+\alpha^2} J (t_c-t)}.
\end{equation}

Predominantly at intermediate skyrmion sizes, according to Eq.\ \eqref{eq:finalODErho}, the DMI term becomes relevant. It depends on the helicity $\varphi_0(t)$ from Eq.\ \eqref{eq:finalODEphi} in a sinusoidal way.

In general, the coupled system of equations \eqref{eq:finalODErho} and \eqref{eq:finalODEphi} cannot be solved analytically. For large skyrmions, as in Eq.\ \eqref{eq:finalODErho}, Eq.\ \eqref{eq:finalODEphi} is dominated by the term proportional to the external magnetic field strength $\tilde{B}$, yielding the precessional solution
\begin{equation}
\label{eq:B_dependence}
\varphi_0 \sim \varphi_{0,0} + \frac{c_5 a^2}{1+\alpha^2} \tilde{B} t,
\end{equation}
where $\varphi_{0,0}$ is the skyrmion helicity at $t=0$. For small skyrmion sizes, on the other hand, the exchange term dominates, which, together with Eq.\ \eqref{eq:squareroot} for this regime, results in a logarithmic time dependence as 
\begin{equation}
\label{eq:ln_precession}
\varphi_0 \sim -\frac{c_8}{2c_4}\ln(t_c-t),
\end{equation}
for the helicity. For times $t\to t_c$ just before the skyrmion collapses, $\varphi_0$ will rapidly increase which, in turn, results in fast oscillations of $\rho_0(t)$ there.

According to Eq.\ \eqref{eq:TriangularEnergy}, the DMI is essential to stabilize skyrmions. It has been shown that the DMI can be controlled externally by electric fields and voltages \cite{SrivastavaBea,KoyamaChiba,KatoHayashi}. In the following, the shrinking of skyrmions is studied for different values of the DMI. If the DMI is quenched to zero at time zero, $\tilde{D}\to 0$, the differential equation for $\rho_0(t)$ in Eq.\ \eqref{eq:finalODErho} decouples from the skyrmions helicity $\varphi_0(t)$. Then the solution of the differential equation reads
\begin{equation}
\label{eq:ODE_solution}
\begin{split}
\rho_0(t)=&\left[\rho_{0,0}^2\exp \left(-\frac{2c_1a^2\alpha}{1+\alpha^2}\tilde{B}t\right) \right. \\
&+ \left. \frac{c_4J}{c_1\tilde{B}}\left( \exp \left(-\frac{2c_1a^2\alpha}{1+\alpha^2}\tilde{B}t \right) -1 \right)\right]^\frac{1}{2}.
\end{split}
\end{equation}
Over time, the radius shrinks and vanishes at the critical time
\begin{equation}
\label{eq:tc}
t_c= \frac{1+\alpha^2}{2c_1a^2\alpha\tilde{B}}\ln\left[ 1+\rho_{0,0}^2\frac{c_1\tilde{B}}{c_4J} \right].
\end{equation}
For times $t\to t_c$ Eq.\ \eqref{eq:ODE_solution} indeed follows the square root behavior of Eq.\ \eqref{eq:squareroot}. Furthermore, re-inserting $\tilde{B}=B/a^2$ reveals that $t_c$ in Eq.\ \eqref{eq:tc} increases with decreasing lattice constant $a$ as $t_c \sim -\ln(a/\rho_{0,0})$, thus stabilizing skyrmions in the continuum limit.

\begin{figure}[t!]
	\centering
	\includegraphics[width=0.47\textwidth]{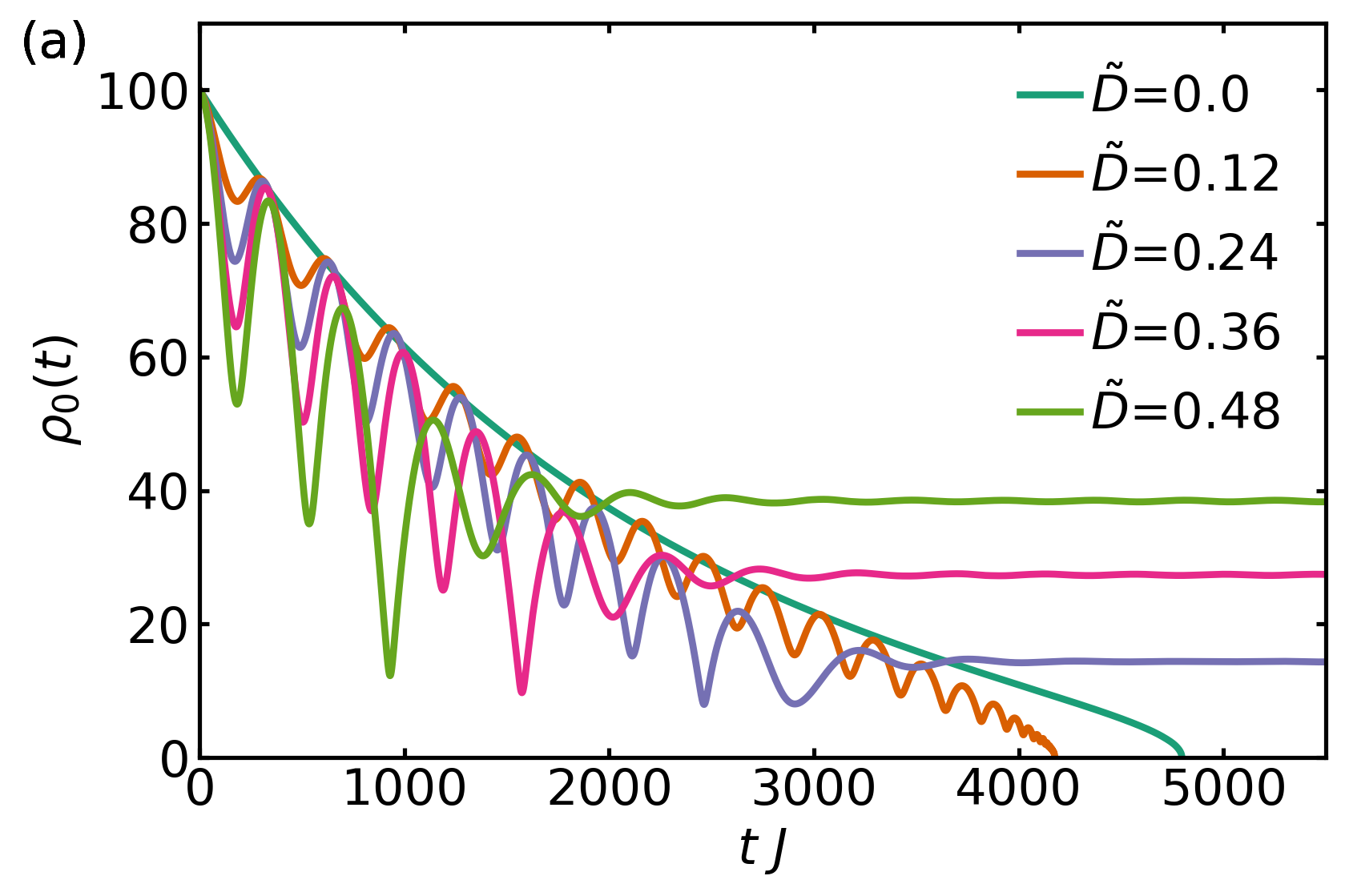}
	\includegraphics[width=0.47\textwidth]{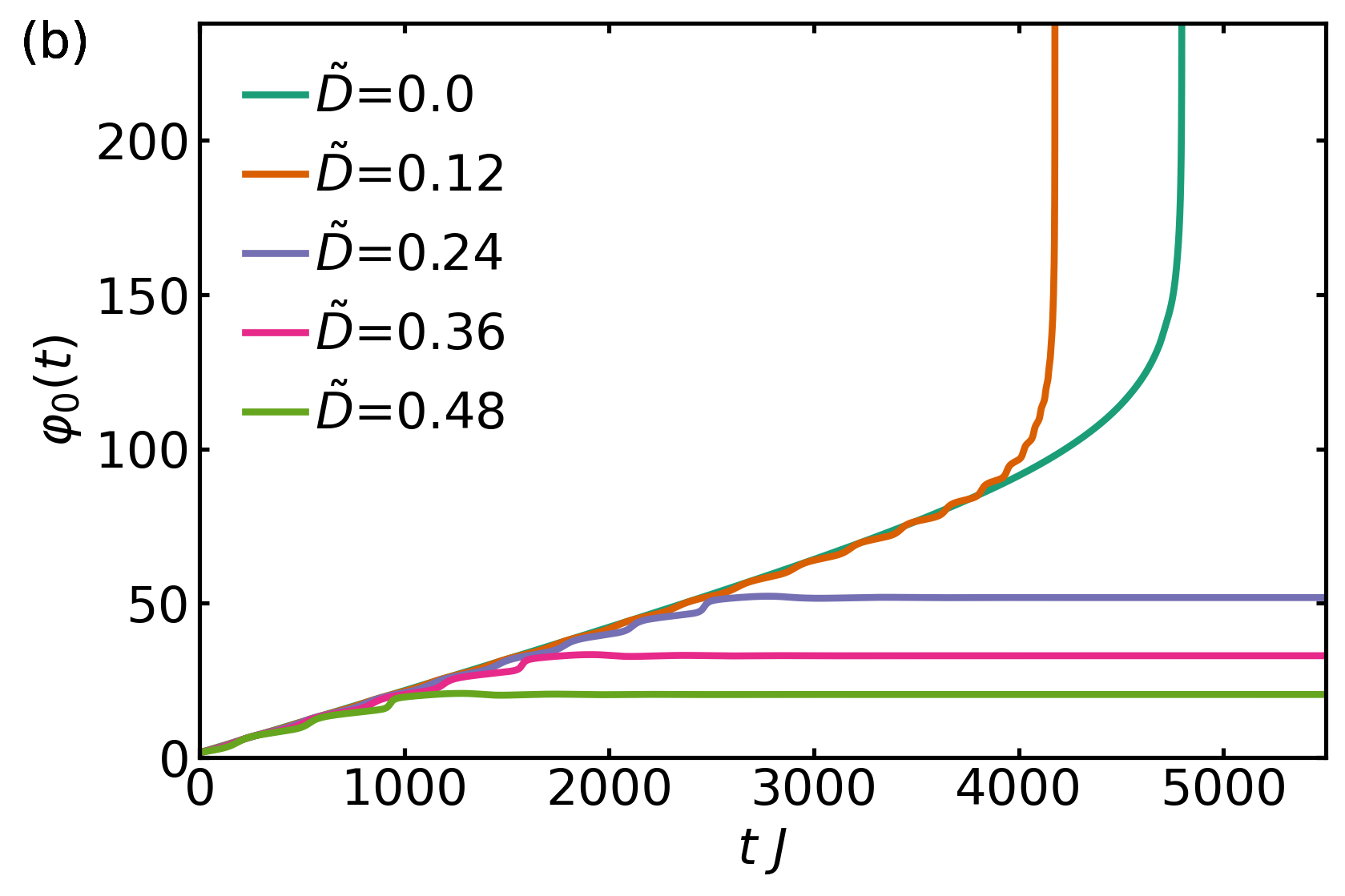}
	\caption{(a) Shrinking dynamics of the radius $\rho_0(t)$ and (b) rotation dynamics of the helicity $\varphi_0(t)$ of a triangular skyrmion for $J=1$, $\tilde{B}=0.02$ and $\alpha=0.1$ for different values of the DMI $\tilde{D}$. Initial conditions where set to $\rho_{0,0}=100a$ and $\varphi_{0,0}=\pi/2$ for $a$ being the lattice constant. The $\tilde{D}=0$ curve in (a) corresponds to the exact analytical solution of the decoupled ODE in \eqref{eq:ODE_solution}, while the other curves for $\tilde{D}>0$ and all curves in (b) are the numerical solutions of the coupled equations Eq.\ \eqref{eq:finalODErho} and Eq.\ \eqref{eq:finalODEphi} for increasing values of the DMI.
		\label{fig:rho0(t)_ODE_tuneD}}
\end{figure}

\begin{figure}[t!]
	\centering
	\includegraphics[width=0.47\textwidth]{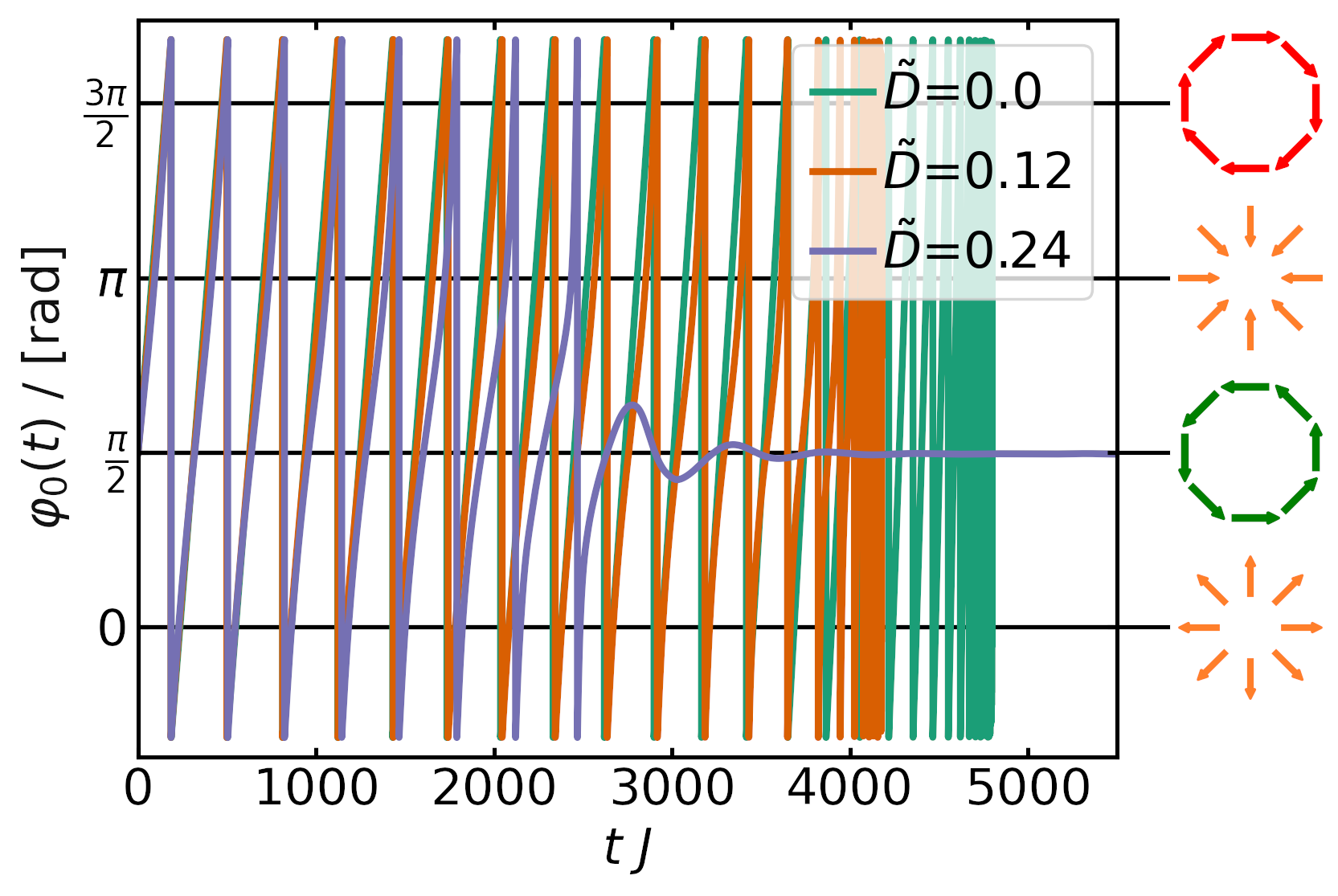}
	\caption{Data of Fig.\ \ref{fig:rho0(t)_ODE_tuneD} (b) wrapped within the interval $[0,2\pi]$. For small DMI values, the skyrmion cycles periodically between the different skyrmion configurations shown on the right side of the plot as pictograms. When the DMI is large enough to stabilize the skyrmion at a finite size, the $\varphi_0(t)$ rotation stops and the skyrmion stays in the Bloch-type configuration.
		\label{fig:phi0(t)_wrapped}}
\end{figure}

The time dependence of the solution for $\tilde{D}=0$ is shown in Fig.\ \ref{fig:rho0(t)_ODE_tuneD} (a) together with the numerical solutions of the coupled equations, Eq.\ \eqref{eq:finalODErho} and Eq.\ \eqref{eq:finalODEphi}, for additional values of $\tilde{D}>0$. The corresponding time evolutions of $\varphi_0(t)$ are presented in (b). Initial conditions where set with a skyrmion radius of $\rho_{0,0}=100a$ and a Bloch-type skyrmion configuration of $\varphi_{0,0}=\pi/2$. The plot demonstrates that, for finite but small DMI, periodic oscillations in $\rho_0(t)$ are induced by the rotation in $\varphi_0(t)$, superimposed with an exponential decay that crosses over to the square-root behaviour at small radii. In the following, these oscillations in the time evolution of the skyrmion radius will be referred to as rotational modes. We observe that an increasing DMI shortens the critical time for collapse. Just before collapse, the frequency of the rotational modes increases with time, cf.\ Eq.\ \eqref{eq:ln_precession}. For sufficiently large DMI, the skyrmion stabilizes at a finite radius. In this high-DMI regime, there is a transition from rotational modes, caused by skyrmion helicity rotation, to a state where $\varphi_0(t)$ halts and the skyrmion stays in the Bloch-type configuration. This is also depicted in Fig.\ \ref{fig:phi0(t)_wrapped}, where the data of Fig.\ \ref{fig:rho0(t)_ODE_tuneD} (b) are wrapped within the interval $[0,2\pi]$; on the right-hand side of the plot the corresponding skyrmion configurations are displayed as pictograms. At this point, the skyrmion exhibits a damped breathing mode around its energetic minimum state. The frequency of this breathing mode differs from that of the rotational mode (see Fig.\ \ref{fig:rho0(t)_ODE_tuneD}). These features are consistent with findings in Ref.\  \cite{McKeeverEverschor-Sitte}, where similar distinctions are made between rotational breathing modes and oscillatory modes.

\section{Comparison with numerical simulations}
In order to verify the approximations used in the above analytical treatment, we use the direct numerical solution of the LLG equation. In our simulations, we consider a two-dimensional square lattice with classical magnetic moments $\bm{n}_i$ at sites $i$ represented by three-dimensional vectors of unit length. The LLG equation in Eq.\ \eqref{eq:LLGlattice} can be brought into a numerically more convenient explicit form
\begin{equation}
\label{eq:LLGexplicit}
\frac{\partial \bm{n}_i}{\partial t}=- \frac{1}{1+\alpha^2} \bm{n}_i \times \bm{B}_i^{\text{eff}} - \frac{\alpha}{1+\alpha^2} \bm{n}_i \times \left(\bm{n}_i \times \bm{B}_i^{\text{eff}} \right) \, , 
\end{equation}
that is used during the simulations. $\bm{B}_i^{\text{eff}}$ is the effective magnetic field at lattice site $i$, defined as
\begin{equation}
\bm{B}_i^{\text{eff}}= -\frac{\partial H}{\partial\bm{n}_i} =J \sum_{j \in \braket{i}} \bm{n}_j + D \sum_{j \in \braket{i}} \left( \bm{n}_j \times \bm{d}_{ij} \right) + B \bm{z}.
\end{equation}
The index $j \in \braket{i}$ runs over the four nearest neighbors of the site $i$ and $\bm{z}$ denotes the unit vector in $z$-direction. The exchange constant $J=1$ is taken as energy unit for different DMI strengths $D$ and external magnetic fields $B$, while Gilbert damping parameter $\alpha$ and the  lattice constant $a$ are also varied. All simulations were performed on a quadratic system, $L_x=L_y=L$. 
To simulate a denser grid, the lattice constant $a$ is decreased, by increasing the number of grid points for a given system size $L$. Thereby the system size $L$, as a length scale, is kept fixed for all simulations. We always ensure to investigate the same skyrmion with the same spatial extension at different grid resolutions, at the same energy. This requires to rescale the system parameters with varying lattice constant according to $J/a^2$ and $D/a$. Initially, a Bloch-type skyrmion is always stabilized for the bulk DMI $D=0.2$ and external field $B=0.02$, until it has reached its skyrmion minimum energy. The so obtained LLG-stationary skyrmion is isotropic in the two-dimensional plane. Then system parameters where changed instantaneously at time $t=0$ which renders the skyrmion configuration instantaneously unstable (e.g., the DMI was turned off ($D=0$)) and the subsequent time evolution can be monitored. The skyrmion starts to shrink until it collapses, dissipating the excess energy into magnon waves. In the following the results of these simulations are discussed. First, we explain how the skyrmion radius $\rho_0$ and the decay time $t_c$ are determined. It is shown how the radius evolves in time for different values of the Gilbert damping, grid density, DMI strength, and the external field. Subsequently, we show how a jump in the topological charge is accompanied by the appearance of a peak in the $n_z(\rho)$-profile, marking the moment of skyrmion decay.

\subsection{Determining the skyrmion radius}
\label{sec:DetSkyrRadius}
To accurately determine the skyrmion radius, we fit the numerical data to the function in Eq.\ \eqref{eq:RadiusFit}, which is given by Wang et al.\ \cite{Wang}. This function is designed to capture the domain wall--like shoulder of width $w$ seen in typical profiles of the polar angle $\Theta(\rho)$ near the radius $\rho_0$. It allows us to fit both parameters, indeed along an arbitrary radial direction, since the skyrmion is isotropic. We have validated that the functional form Eq.\ \eqref{eq:RadiusFit} gives good fitting results of the skyrmion profile.

\subsection{Determining the decay time}
\label{sec:DetDecTime}
\begin{figure}[t!]
	\centering
	\includegraphics[width=0.47\textwidth]{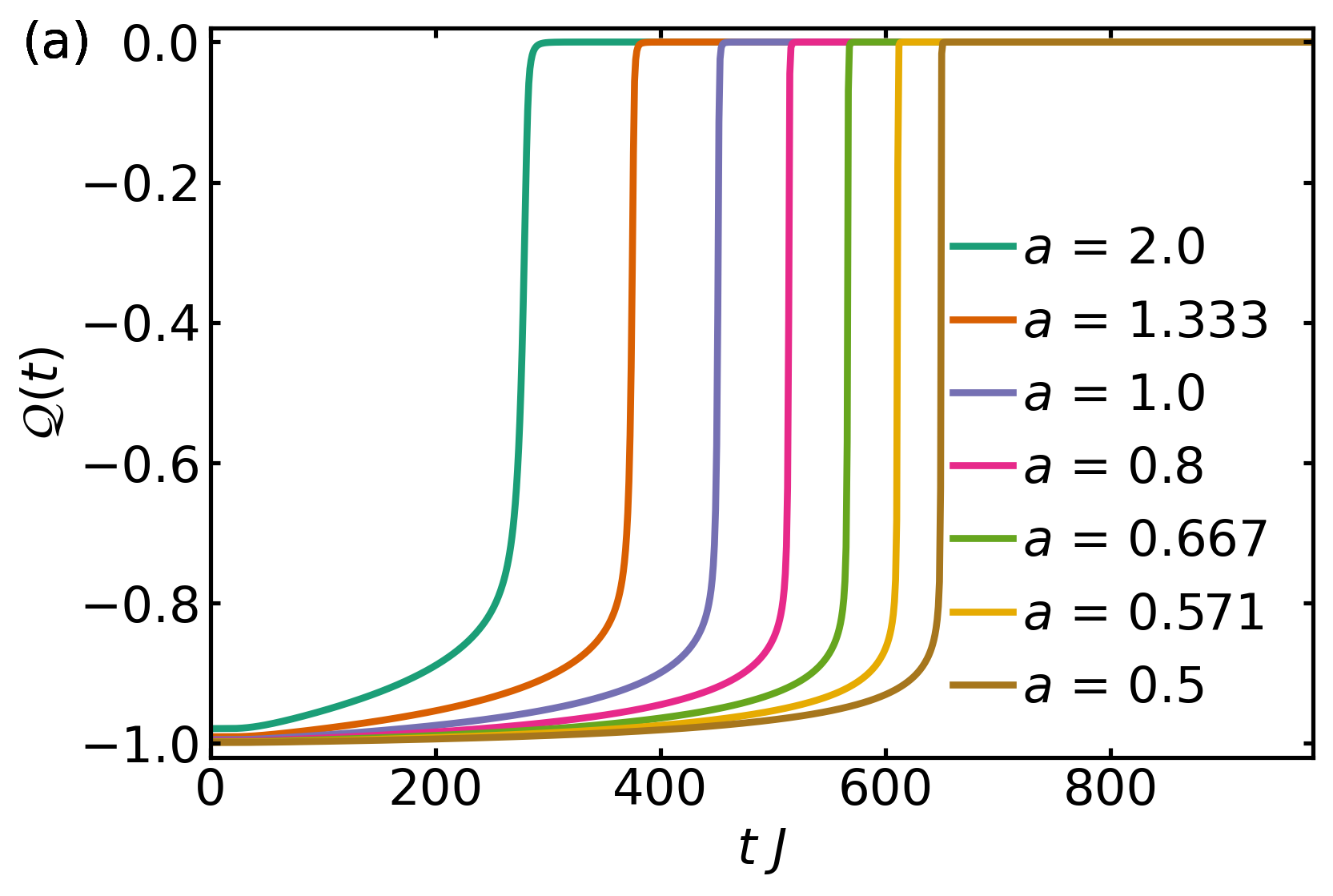}
	\includegraphics[width=0.47\textwidth]{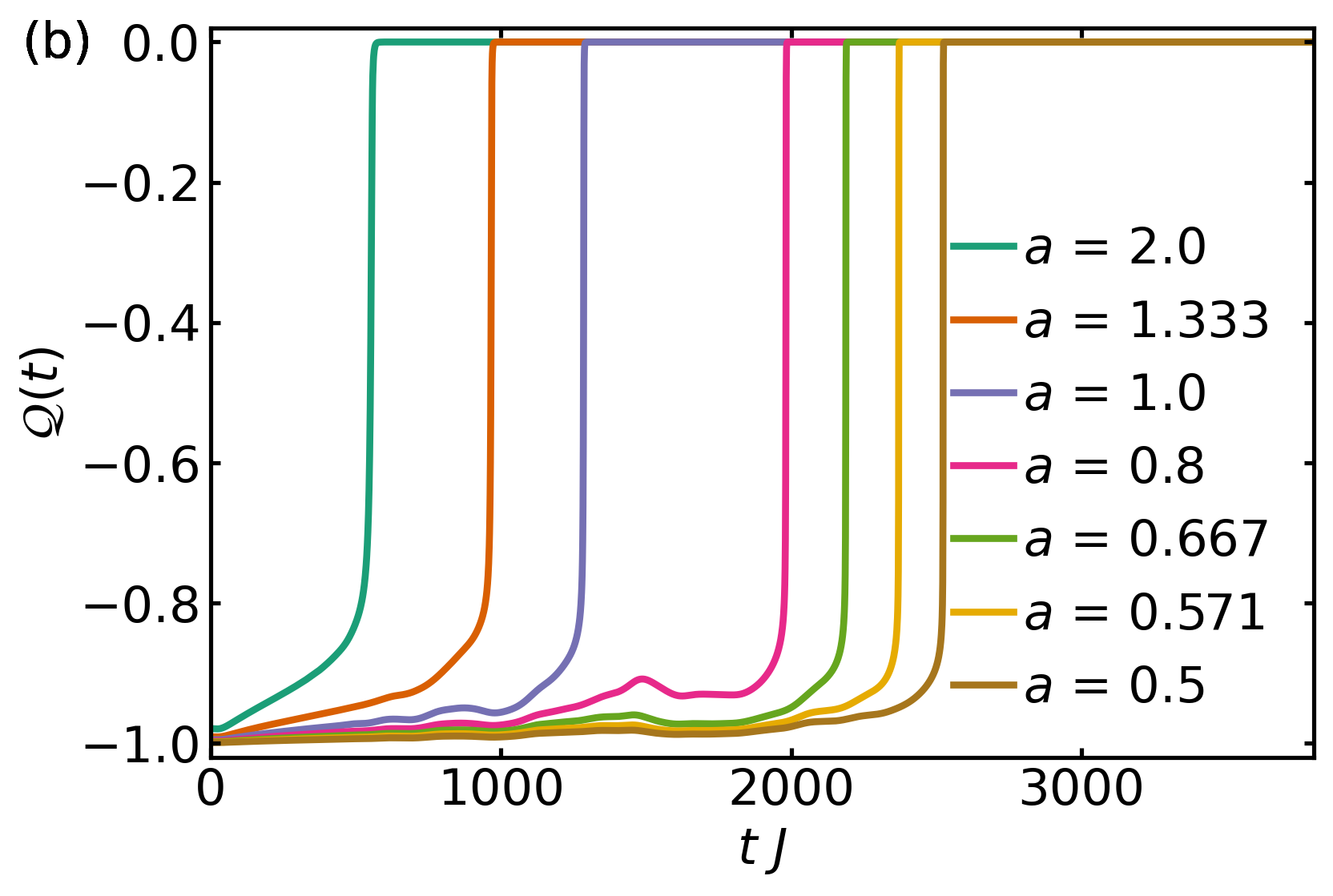}
	\caption{Time evolution of the total topological charge of the skyrmion for different lattice constants with a Gilbert damping of $\alpha=0.1$ (a) and $0.01$ (b), both at $D=0$ and $B=0.02$.
		\label{fig:Qtot}}
\end{figure}

To determine the point in time, $t_c$, that marks the skyrmion decay, the topological charge of the skyrmion is calculated according to Eq.\ \eqref{eq:topCharge}, at each time step. This leads to Fig.\ \ref{fig:Qtot} where the time evolution of the total topological charge $\mathcal{Q}(t)$ is shown for $\alpha = 0.1$ (a) and $\alpha = 0.01$ (b), using different lattice constants $a$. We find that the smaller the lattice constant is the later the jump in the topological charge appears, from $\mathcal{Q} \approx -1$ (skyrmion exists) to $\mathcal{Q} \approx 0$ (only a ferromagnetic phase exists). This already indicates an increasing skyrmion decay time $t_c$ with decreasing lattice constant. Also, the general impact of the Gilbert damping on the decay time can be identified. With increasing Gilbert damping the skyrmion decay time $t_c$ decreases for a fixed lattice constant. Both observations are consistent with Eq.\ \eqref{eq:tc}.

\begin{figure}[t!]
	\centering
	\includegraphics[width=0.47\textwidth]{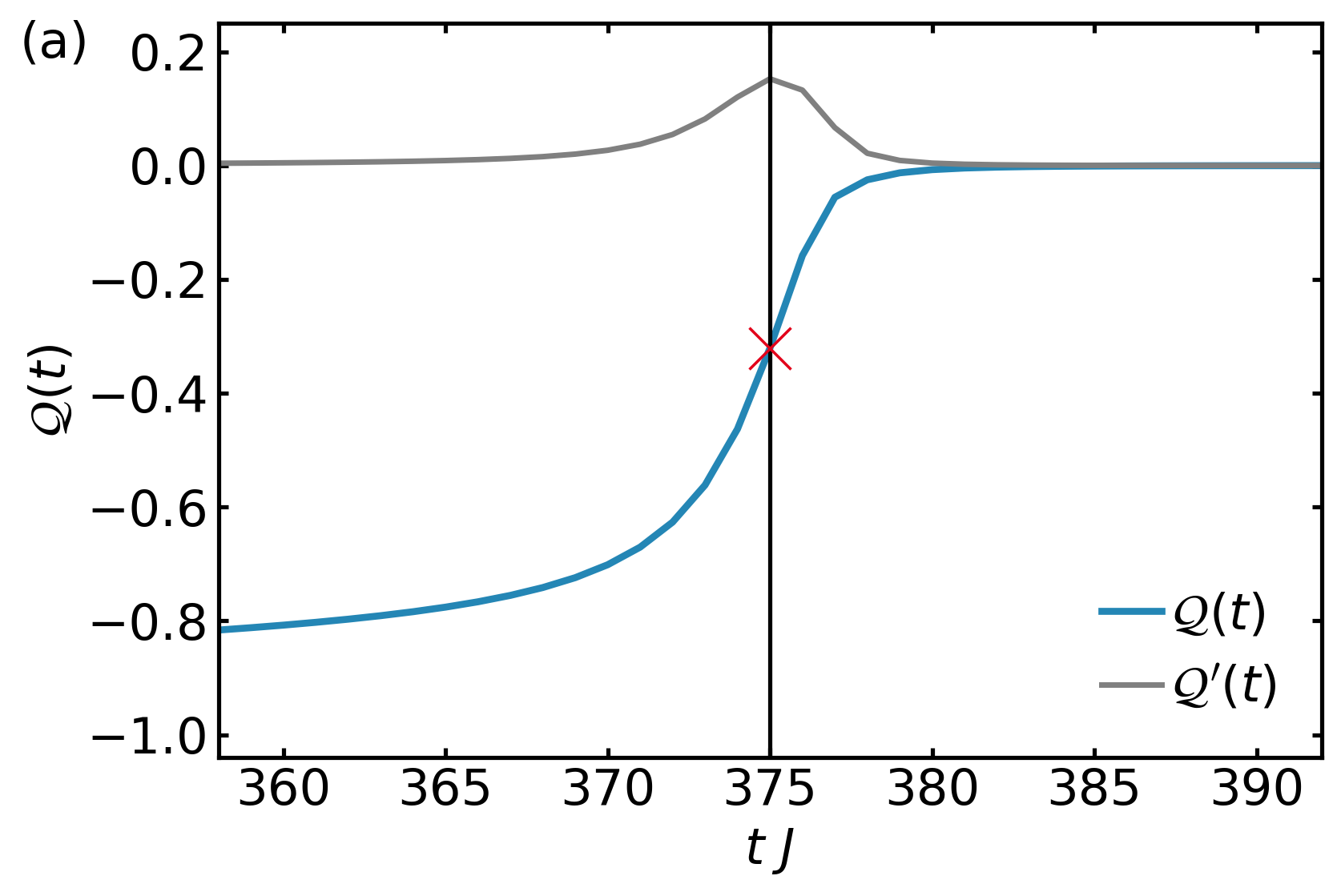}
	\includegraphics[width=0.47\textwidth]{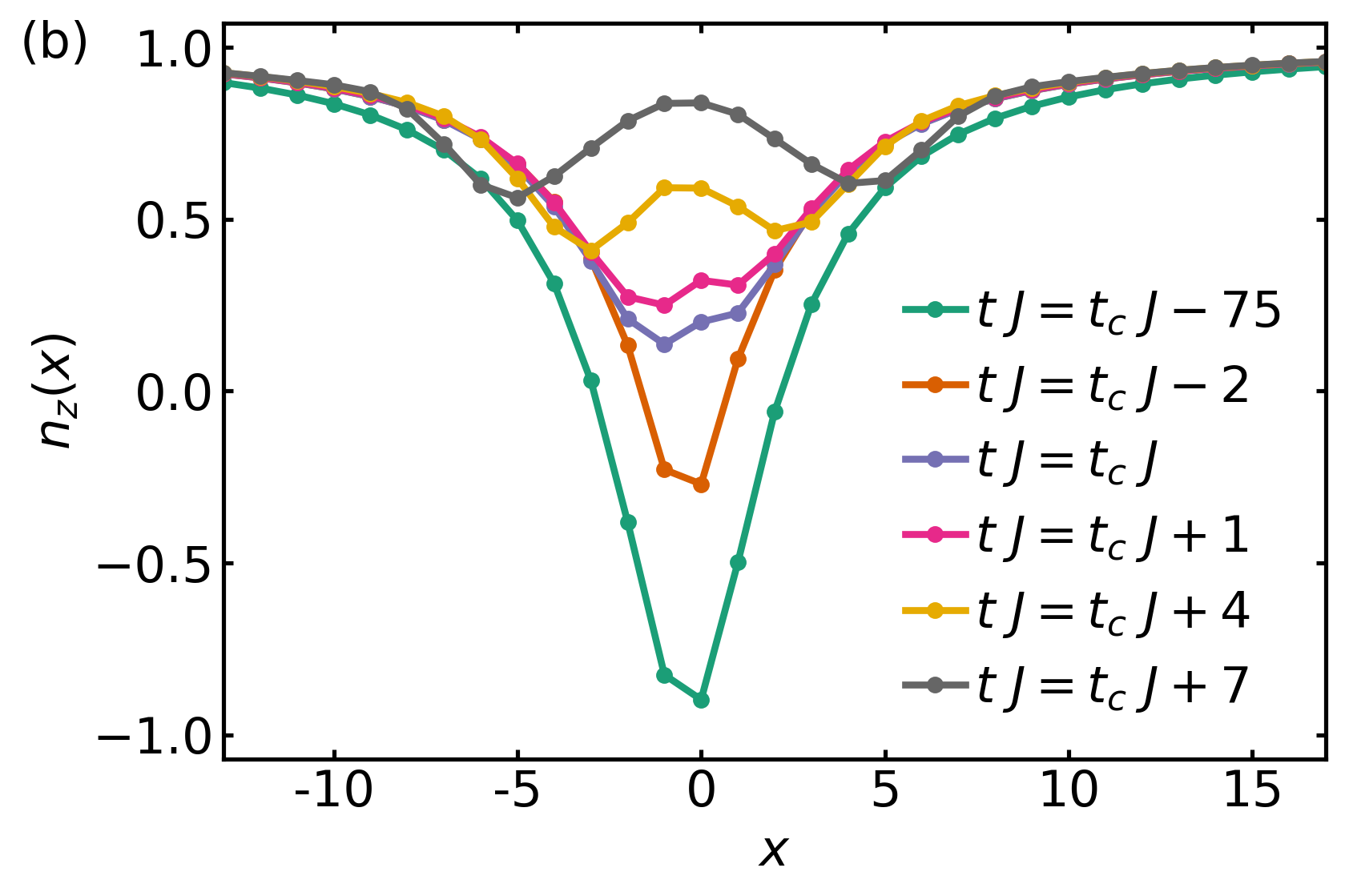}
	\caption{(a) Zoom of $\mathcal{Q}(t)$ (blue line) around the time of the skyrmion collapse for $a=1$ and $\alpha=0.01$. The grey line depicts the numerical derivative $\mathcal{Q}'(t)$ and the red cross indicates the point of time when $\mathcal{Q}'(t)$ is maximal. (b) Radial profile through the center of the skyrmion for different time points during the simulation. The blue line corresponds to the time step marked by the red cross in (a), the green and orange lines represent the profiles for earlier times and the magenta, yellow and grey lines for later times. The latter show an opening of the $n_z(x)$-profile, marking the decay of the skyrmion.
		\label{fig:DecayTime}}
\end{figure}

In Fig.\ \ref{fig:DecayTime}, we show an enlarged zoom to the decaying skyrmion. Fig.\ \ref{fig:DecayTime} (a) shows a zoom of $\mathcal{Q}(t)$ (blue line) obtained from the simulation for $a=1$ and $\alpha=0.01$ around the point of time when the skyrmion collapse happens (red cross in Fig.\ \ref{fig:DecayTime} (a)). The grey line shows the numerical time derivative $\mathcal{Q}'(t)$  of the topological charge to find the inflection point in the time evolution of $\mathcal{Q}$. In Fig.\ \ref{fig:DecayTime} (b), we investigate the radial profile of the $z$-component of the magnetization field at different times. There, the blue line shows the radial profile through the skyrmion center at the point of time where the derivative has its maximum. The green and orange lines show the profile for earlier times during the simulation and the magenta, yellow and grey lines for later times. It is evident that the so found time point is the last one before the $n_z(x)$-profile becomes non-monotonous at the origin, marking the skyrmions decay time $t_c$. This behavior is reproducible for all the other simulations, making it a robust procedure to determine $t_c$.

\begin{figure}[t!]
	\centering
	\includegraphics[width=0.47\textwidth]{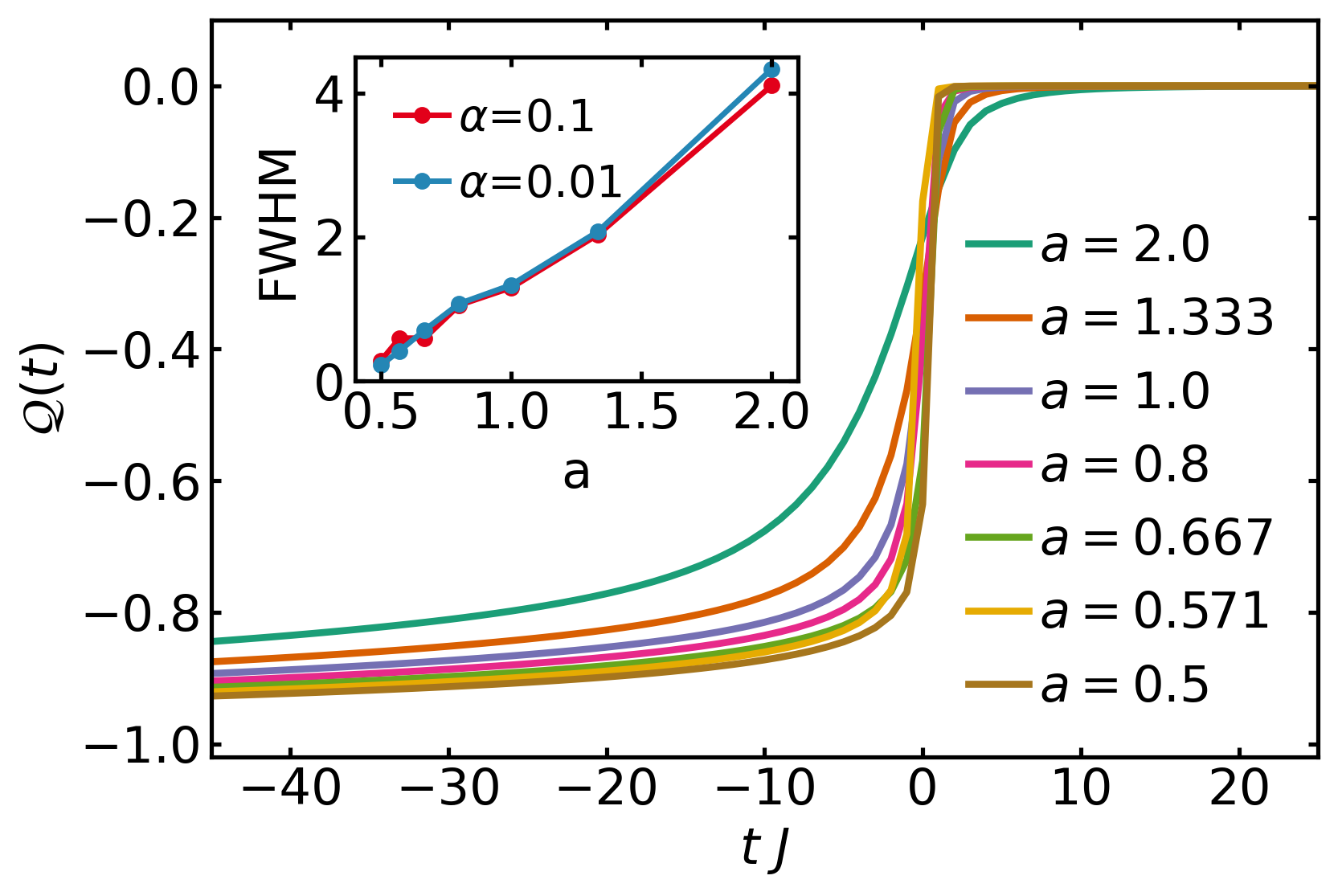}
	\caption{Time-dependent total topological charge (same data as in  Fig.\ \ref{fig:Qtot} for $\alpha=0.1$, but shifted in time such that the collapse in the topological charge is aligned) for different values of the lattice constant. With decreasing lattice constant, the jump becomes steeper. To quantify this, the inset shows the FWHM of Lorentzians fitted to the maxima of the time derivative $\mathcal{Q}'(t)$ in dependence of the lattice constant for $\alpha=0.1$ and $0.01$.
		\label{fig:Qtot_aligned}}
\end{figure}

Next, we study the dependence of the skyrmion collapse on the lattice constant. In Fig.\ \ref{fig:Qtot_aligned}, we show the time-dependent total topological charge for different lattice constants $a$ for $\alpha=0.1$, artificially shifted to the same critical collapse time $t_c$ to emphasize how the collapse becomes steeper in time for smaller lattice constants. For a given $a$, each collapse time is thereby identified from the maximum of the fitted Lorentzian to the derivative of $\mathcal{Q}'(t)$. In turn, the steepness of the jump in $\mathcal{Q}(t)$ with decreasing lattice constant is quantified by their corresponding full width at half maximum (FWHM). The resulting dependence of the FWHM on the different lattice constants is shown for $\alpha=0.1$ and $\alpha=0.01$, respectively, as inset of Fig.\ \ref{fig:Qtot_aligned}. It can clearly be seen that the FWHM decreases for decreasing lattice constants confirming the statement from above. On the other hand, the FWHM for the different damping strengths does not differ significantly, indicating that the decay process itself does not essentially depend on the Gilbert damping $\alpha$. In turn, $\alpha$ only influences $t_c$, while the lattice constant has an impact on both, the decay time $t_c$ and on how fast the collapse itself happens.

\subsection{Time evolution of the skyrmion radius for $D=0$}
\begin{figure}[t!]
	\centering
	\includegraphics[width=0.47\textwidth]{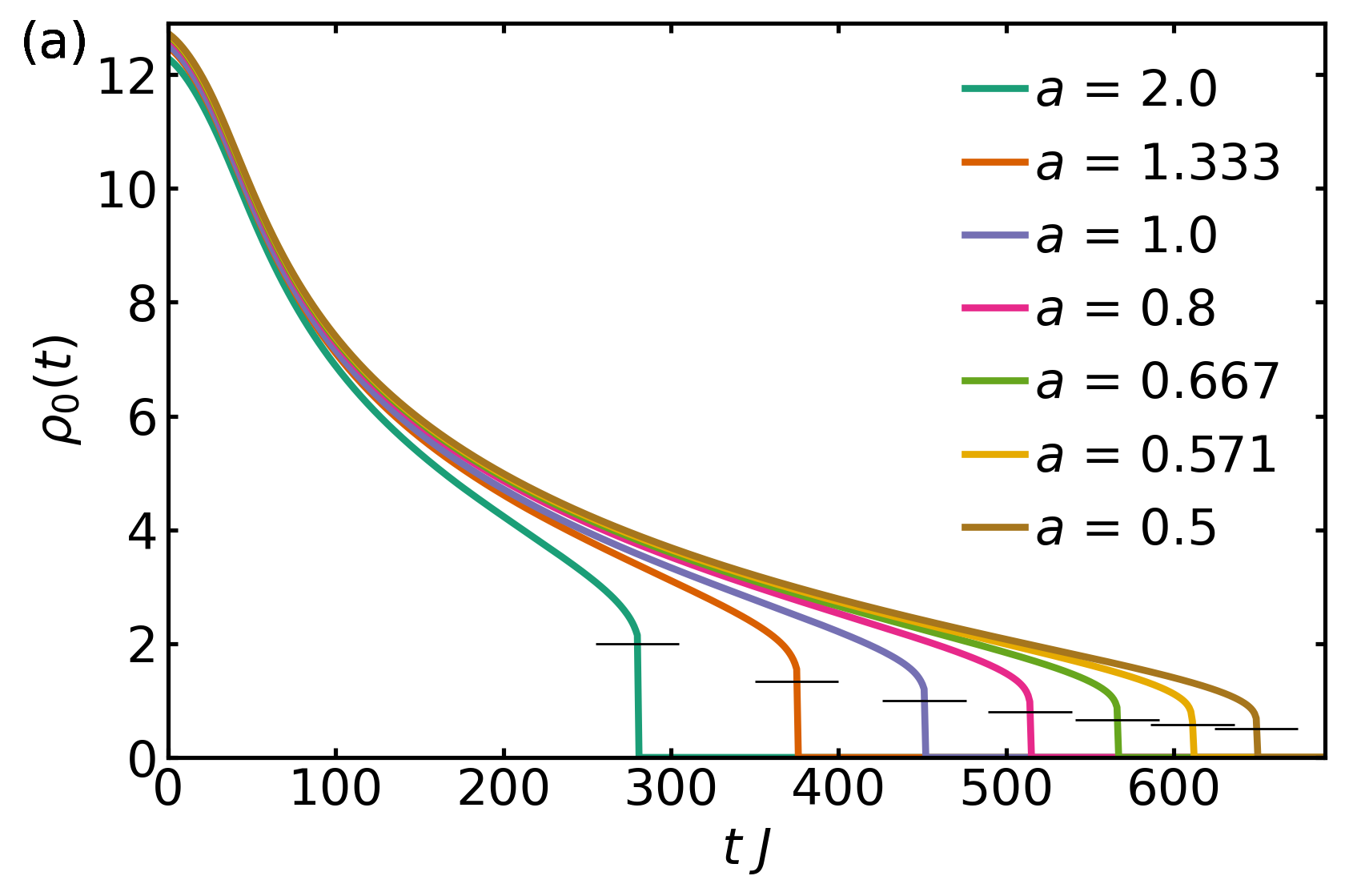}
	\includegraphics[width=0.47\textwidth]{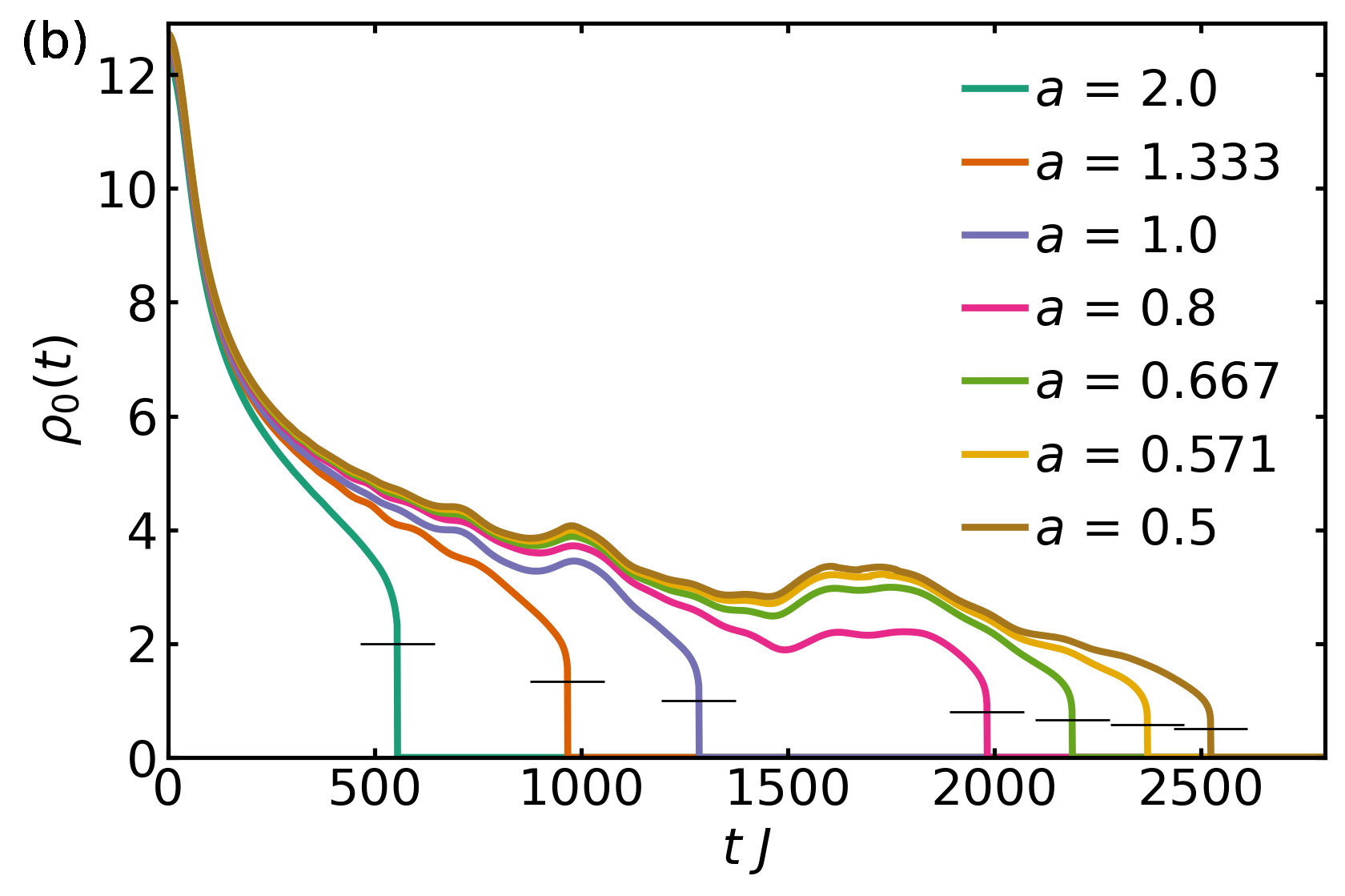}
	\caption{Time evolution of the skyrmion radius for different lattice constants and for a Gilbert damping of $\alpha=0.1$ (a) and $\alpha=0.01$ (b), both at $D=0$ and $B=0.02$. The horizontal black lines mark the lattice constant used for the corresponding simulation. For low damping, complicated vibrations make the skyrmion radius follow some non-periodic breathing-like modes. These vibrations are suppressed for higher damping, making the skyrmion shrinking to follow a smooth curve.
		\label{fig:rho0_num}}
\end{figure}

Using the procedure described in Secs.\ \ref{sec:DetSkyrRadius} and \ref{sec:DetDecTime}, the time evolution of the skyrmion radius $\rho_0(t)$ has been determined for $D=0$ and is shown in Fig.\ \ref{fig:rho0_num}. The evolution of the radii for different lattice constants is shown in Fig.\ \ref{fig:rho0_num} (a) for $\alpha=0.1$ and in Fig.\ \ref{fig:rho0_num} (b) for $\alpha=0.01$. The horizontal black lines in the plots mark the lattice constant used for the corresponding simulation. The last time point where the skyrmion still exists is always the last simulated time step when its radius is larger than the lattice constant. So, the time $t_c$, when the skyrmion decays, is reached when its radius would become smaller than the lattice constant.

In all simulations the skyrmion starts with a Bloch-type configuration. However, as it shrinks, we find that the magnetic moments collectively rotate through the skyrmion configurations from Bloch to anti-N\'eel to anti-Bloch and to N\'eel configurations before reverting back to the Bloch form, as predicted by our model Eq.\ \eqref{eq:finalODEphi} for $\tilde{D}=0$. Moreover, in the case of smaller Gilbert damping we find that the skyrmion radius does not decrease monotonically but alternately shrinks and growths before it finally vanishes. Damping then seems not strong enough to suppress collective internal vibrational modes in the time evolution of the magnetic moments, resulting in these breathing-like modes. Stronger damping suppresses this motion so that the skyrmion shrinks monotonically in time, such as seen in Fig.\ \ref{fig:rho0(t)_ODE_tuneD} (a) for $\tilde{D}=0$.

\begin{figure}[t!]
	\centering
	\includegraphics[width=0.47\textwidth]{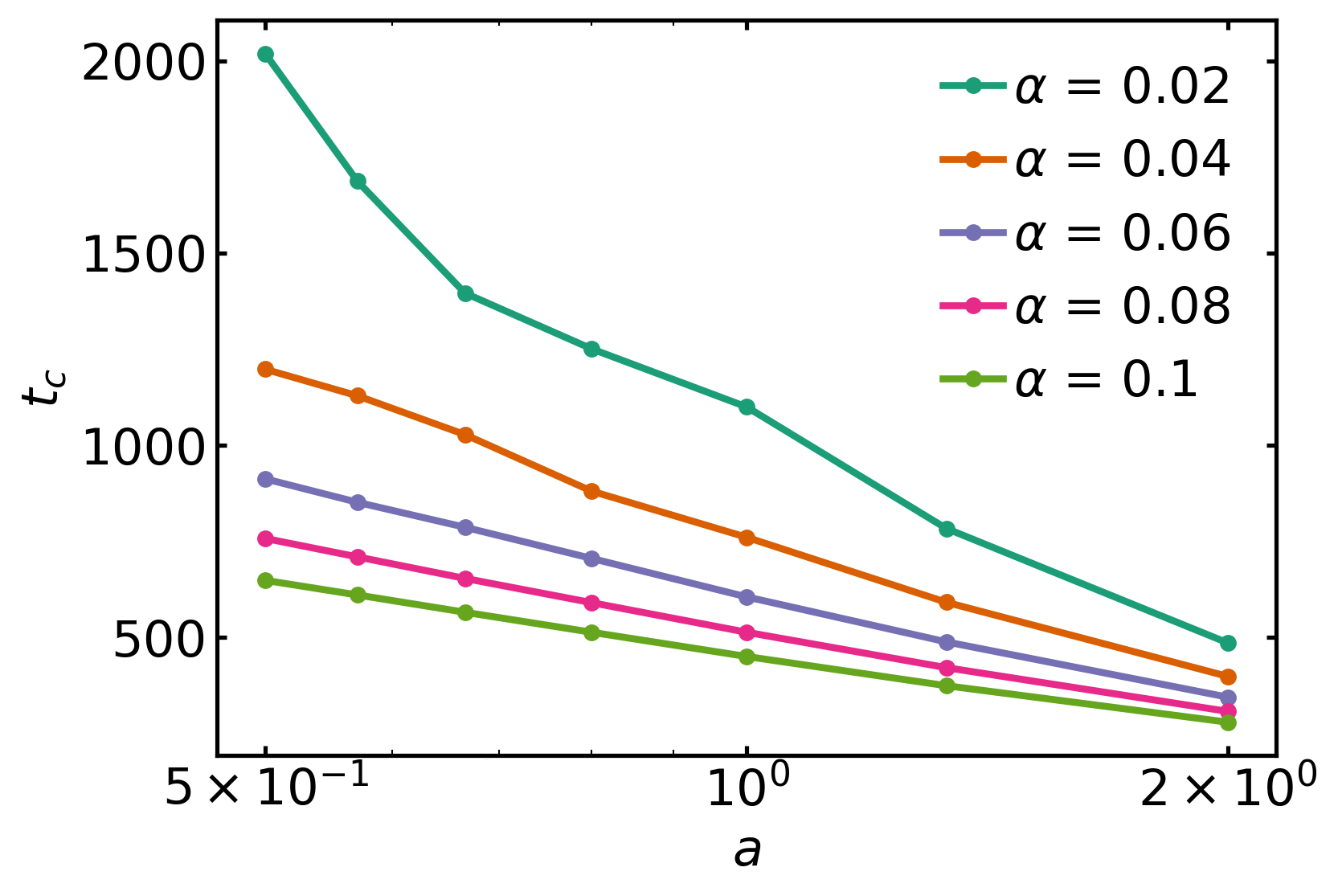}
	\caption{Skyrmion decay times $t_c$ as a function of the lattice constant $a$ for different values of the Gilbert damping $\alpha$ at $D=0$ and $B=0.02$. For small damping, internal breathing modes of the skyrmion appear in the time evolution of the skyrmion radius. These modes are suppressed at stronger damping and $t_c$ shows a purely logarithmic dependence on the lattice constant. For a fixed lattice constant $t_c$ occurs later the smaller the damping is.
		\label{fig:tc_alpha}}
\end{figure}

To investigate the dependence of $t_c$ on the Gilbert damping, additional simulations have been performed for more values of  $\alpha$. In Fig.\ \ref{fig:tc_alpha} the decay time $t_c$ is plotted for different values of $\alpha$ in dependence of the lattice constant in a logarithmic plot. This shows that for small values of the damping $\alpha\leq 0.05$ breathing modes appear that are suppressed at larger values of the damping for $\alpha\geq 0.06$. The linearity of the plots in this regime of larger damping suggests an in general logarithmic dependence of $t_c$ on the lattice constant, in accordance with Eq.\ \eqref{eq:tc}. Due to the breathing at lower damping the skyrmion radius may reach the critical radius, which is of the order of the lattice constant, only later. This explains the deviation from the logarithmic behavior in this regime. Further it is seen that for increasing $\alpha$ the decay time $t_c$ decreases at a given value of $a$.

\begin{figure}[t!]
	\centering
	\includegraphics[width=0.47\textwidth]{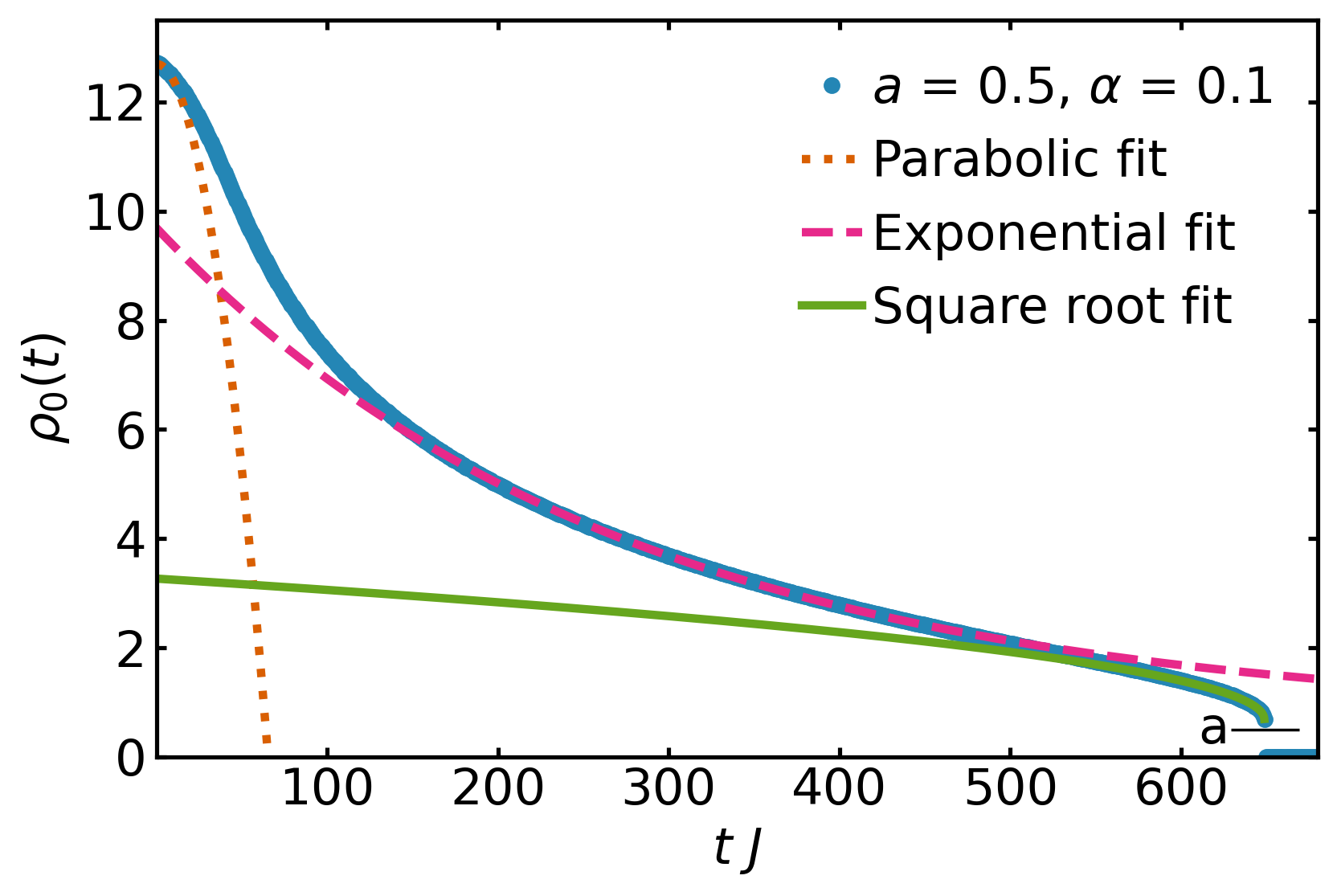}
	\caption{Time evolution of a skyrmion decay. For short times, the system reacts delayed and evolves in quadratic order (orange dotted parabolic fit) in response to the sudden change of the DMI strength. Afterwards, the skyrmion shrinks according to an exponential law (magenta dashed line), that finally crosses over into a square root-like behavior as predicted by the triangular approximation, Eq.\ \eqref{eq:squareroot}, before its radius reaches the value of the lattice constant and the skyrmion collapses.
		\label{fig:FitSkyrmionDecay_num}}
\end{figure}

To further illuminate the behavior of the skyrmion shrinking, Fig.\ \ref{fig:FitSkyrmionDecay_num} shows the skyrmion decay at $B=0.02$, $D=0$ and $\alpha=0.1$ for a lattice constant of $a=0.5$ as a blue line. At $t=0$, the DMI is abruptly turned off and, as usual for quenches, the system reacts delayed to the sudden change of system parameters. The dynamics 
initially follows a quadratic time dependence --- the decay in the beginning can be fitted by a parabola (orange dotted line) --- until the system has adjusted to the new situation. Then, the skyrmion shrinks, following an exponential law (magenta dashed line). Later it crosses over into a square root-like behavior (green line) for small skyrmion radii, just as our theoretical model predicts, before reaching values of the order of the lattice constant, at which point the skyrmion collapses.

\subsection{Time evolution of the skyrmion radius for finite DMI}
\begin{figure}[t!]
	\centering
	\includegraphics[width=0.47\textwidth]{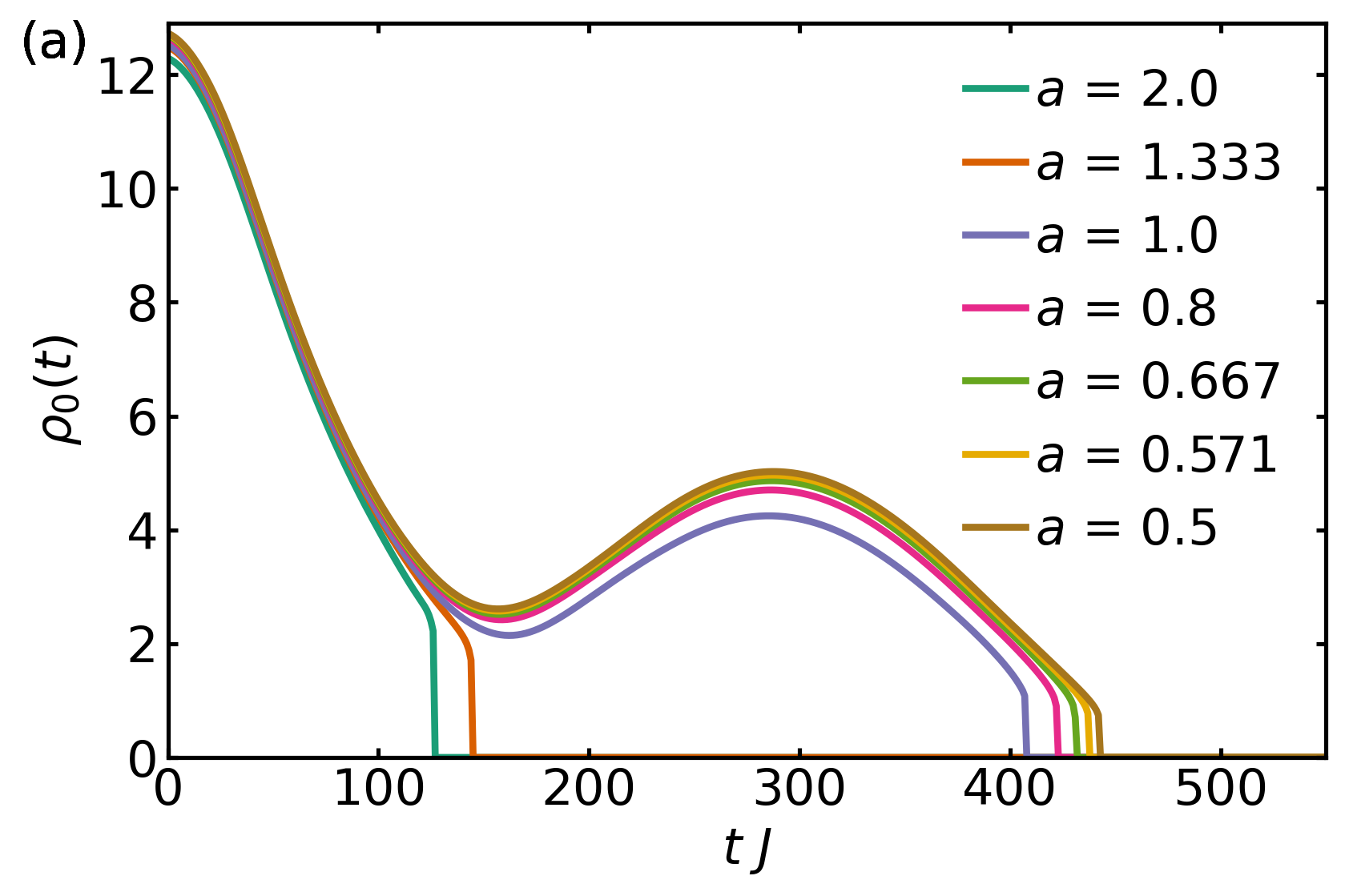}
	\includegraphics[width=0.47\textwidth]{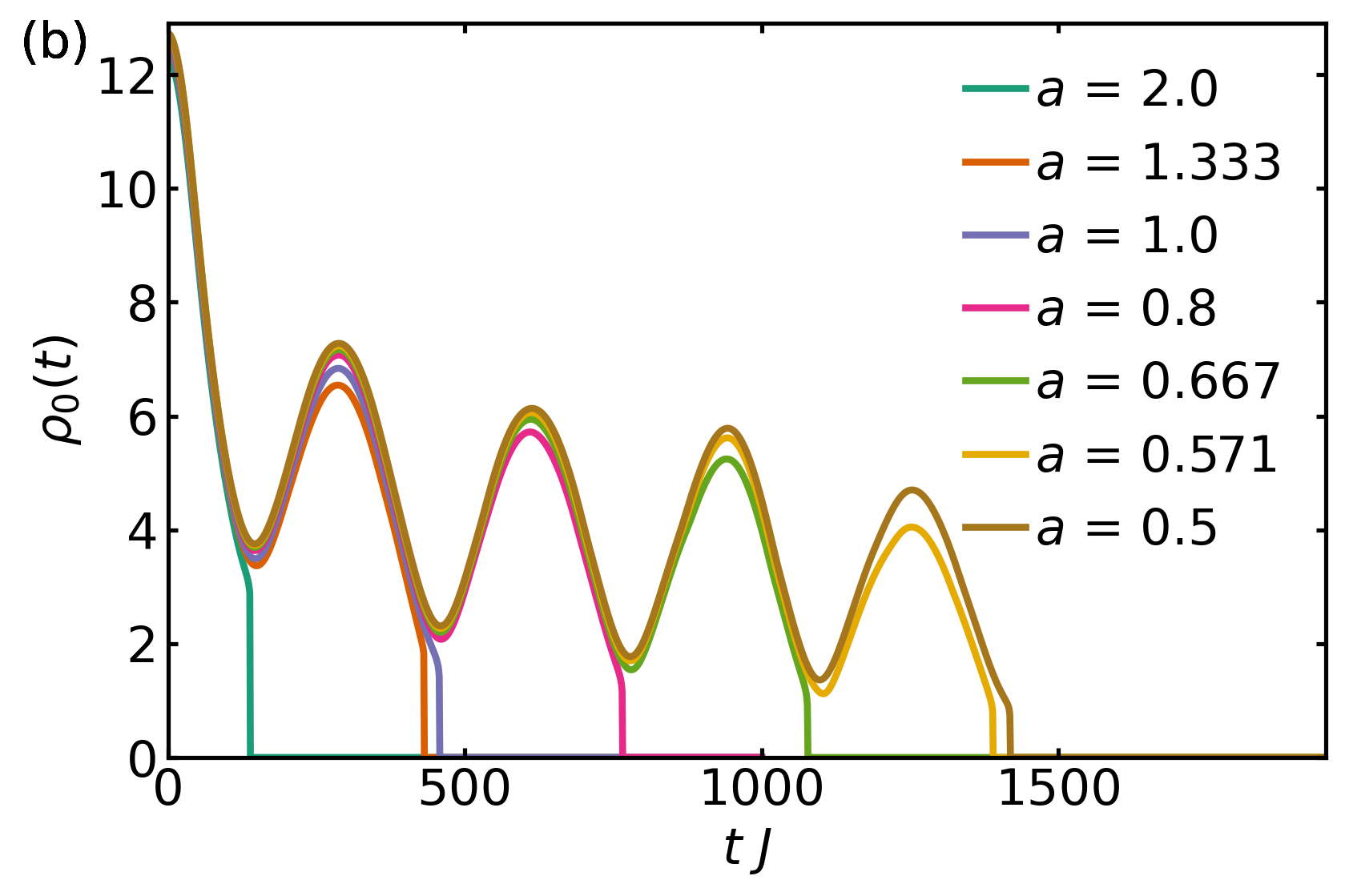}
	\caption{Time evolution of the skyrmion radius for different lattice constants and for $\alpha=0.1$ (a) and $\alpha=0.01$ (b) for $D=0.04$ and $B=0.02$. In both cases rotational modes appear that are periodic in nature as predicted by the coupled differential equations. The period length is the same in both plots, suggesting that it is independent of the damping. However, the larger damping leads to a faster shrinking of the skyrmion such that the skyrmion decay time $t_c$ is reached earlier.
		\label{fig:rho0_D0.04}}
\end{figure}

Next, we present the simulation results for finite DMI. In this parameter regime, periodic rotational modes become apparent, as predicted by our model for a finite DMI, that is not large enough to stabilize the skyrmion. The skyrmion keeps alternating between Bloch- and N\'eel-type configurations throughout its decrease. The time evolution of the skyrmion radius for different lattice constants and for $D=0.04$  and $B=0.02$ are shown in Fig.\ \ref{fig:rho0_D0.04} for $\alpha=0.1$ (a) and $0.01$ (b). Rotational modes occur periodically and the decay happens in the moment when the minimum in the amplitude becomes smaller than the lattice constant. The period length of the rotational mode is not affected by the strength of the damping. Also, larger damping, as expected, does not suppress rotational modes, but leads in general to a faster shrinking of the skyrmion radius so that $t_c$ is reached earlier.

\begin{figure}[t!]
	\centering
	\includegraphics[width=0.47\textwidth]{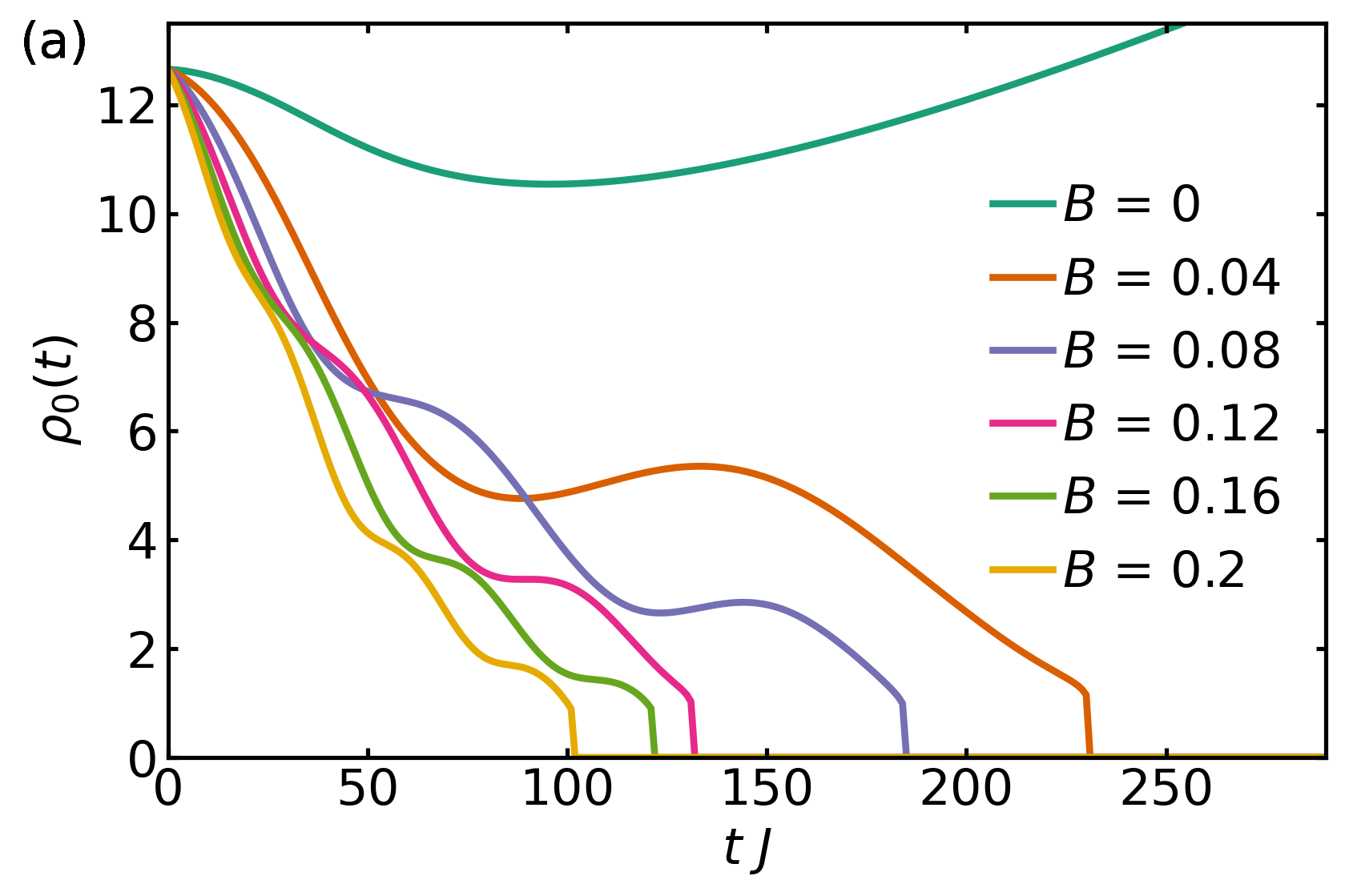}
	\includegraphics[width=0.47\textwidth]{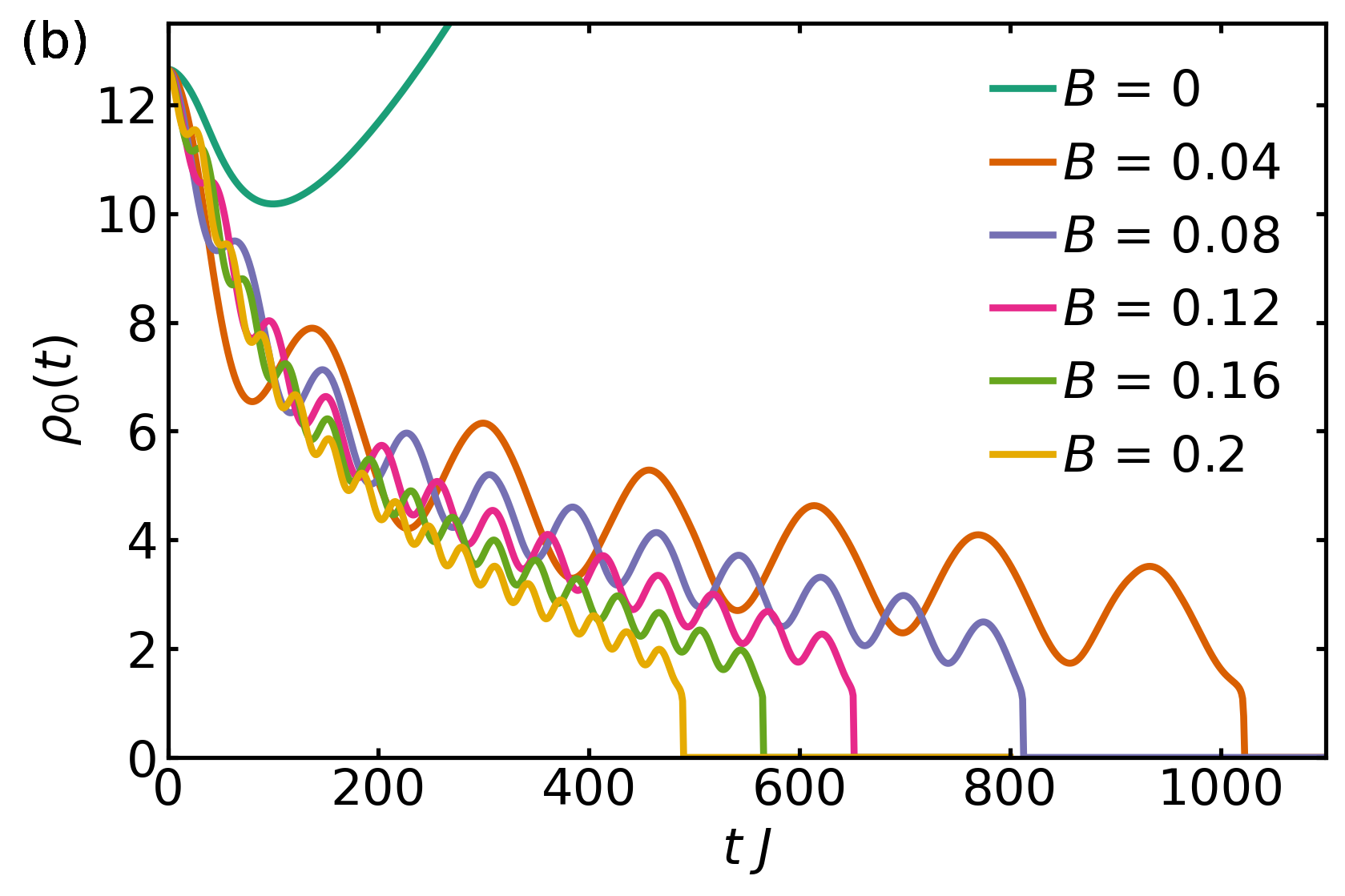}
	\caption{Time evolution of the skyrmion radius for varying values of the external field $B$ and for a Gilbert damping of $\alpha=0.1$ (a) and $\alpha=0.01$ (b) at a DMI of $D=0.04$ for a lattice constant of $a=0.8$. With increasing field strengths, the amplitude of the rotational modes decreases, while their frequency increases, leading to decreasing decay times $t_c$.
		\label{fig:rho0_D0.04_num_tuneB}}
\end{figure}

% \subsection{Decay at varying external fields}
Fig.\ \ref{fig:rho0_D0.04_num_tuneB} depicts the time evolution of the skyrmion radius for different values of the external magnetic field for $D=0.04$ and $\alpha=0.1$ (a) and $\alpha=0.01$ (b), with a lattice constant $a=0.8$. With increasing external magnetic field, the amplitude of the rotational mode decreases, while the frequency increases, following Eq.\ \eqref{eq:B_dependence}, which leads in general to shorter decay times $t_c$. For $B=0$, the skyrmion size $\rho_0$ increases continuously. This behavior is comparable to the Belavin-Polyakov case \cite{BelavinPolyakov}. The Belavin-Polyakov skyrmion exhibits scale invariance, meaning there is no intrinsic length scale that energetically favors a specific skyrmion size. In the absence of external constraints or terms that break this invariance, such as an external magnetic field, the skyrmion can expand indefinitely, similar to what is observed here. The system lacks an easy axis and the skyrmion maintains the Bloch-type configuration with no rotations of its helicity.

\begin{figure}[t!]
	\centering
	\includegraphics[width=0.47\textwidth]{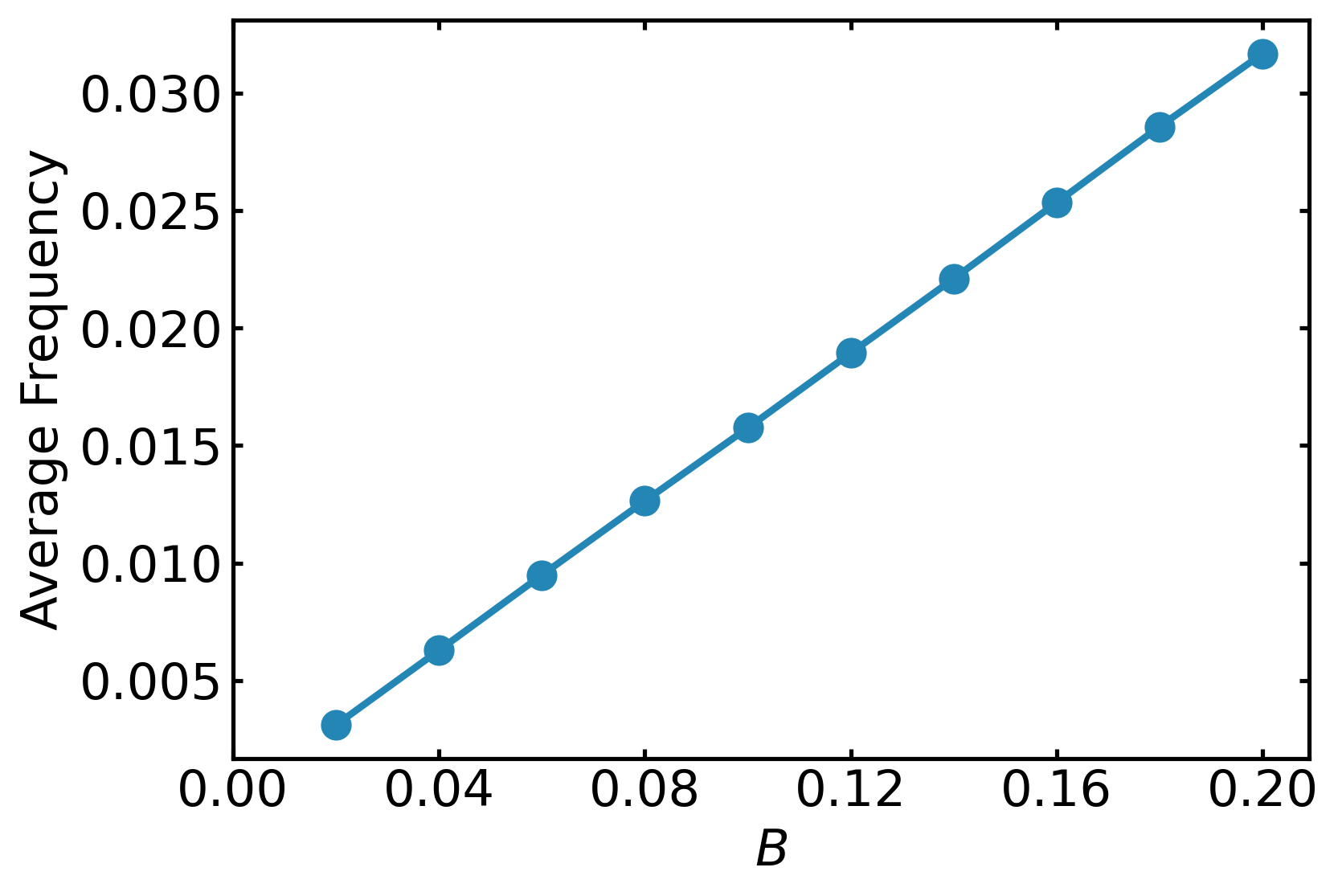}
	\caption{The frequency of the dominant skyrmion rotational mode shows a linear dependence on the external magnetic field strength $B$ as predicted by Eq.\ \eqref{eq:B_dependence}.
	\label{fig:AverageFrequencyVsB}}
\end{figure}

The rotational mode can be further analyzed by the Fourier transform of the skyrmion radius, revealing that a single mode is dominant. Thus, the shrinking follows an exponential decay superimposed with a damped harmonic function, until the radius reaches the value of the lattice constant when the skyrmion collapses. In Fig.\ \ref{fig:AverageFrequencyVsB}, the frequency of the dominant rotational mode for $\alpha=0.01$ in dependence of the external magnetic field strength $B$ is plotted, revealing a linear dependence of the oscillation frequency on the field strength, as the analytic model predicts for large radii. Increasing oscillation frequencies at small radii are not found in the simulations. This is probably due to the fact, that the collapse when reaching the size of the lattice constant happens before entering the really small radii regime.

%\subsection{Fitting the periodic breathing modes at finite DMI}

\begin{figure}[t!]
	\centering
	\includegraphics[width=0.47\textwidth]{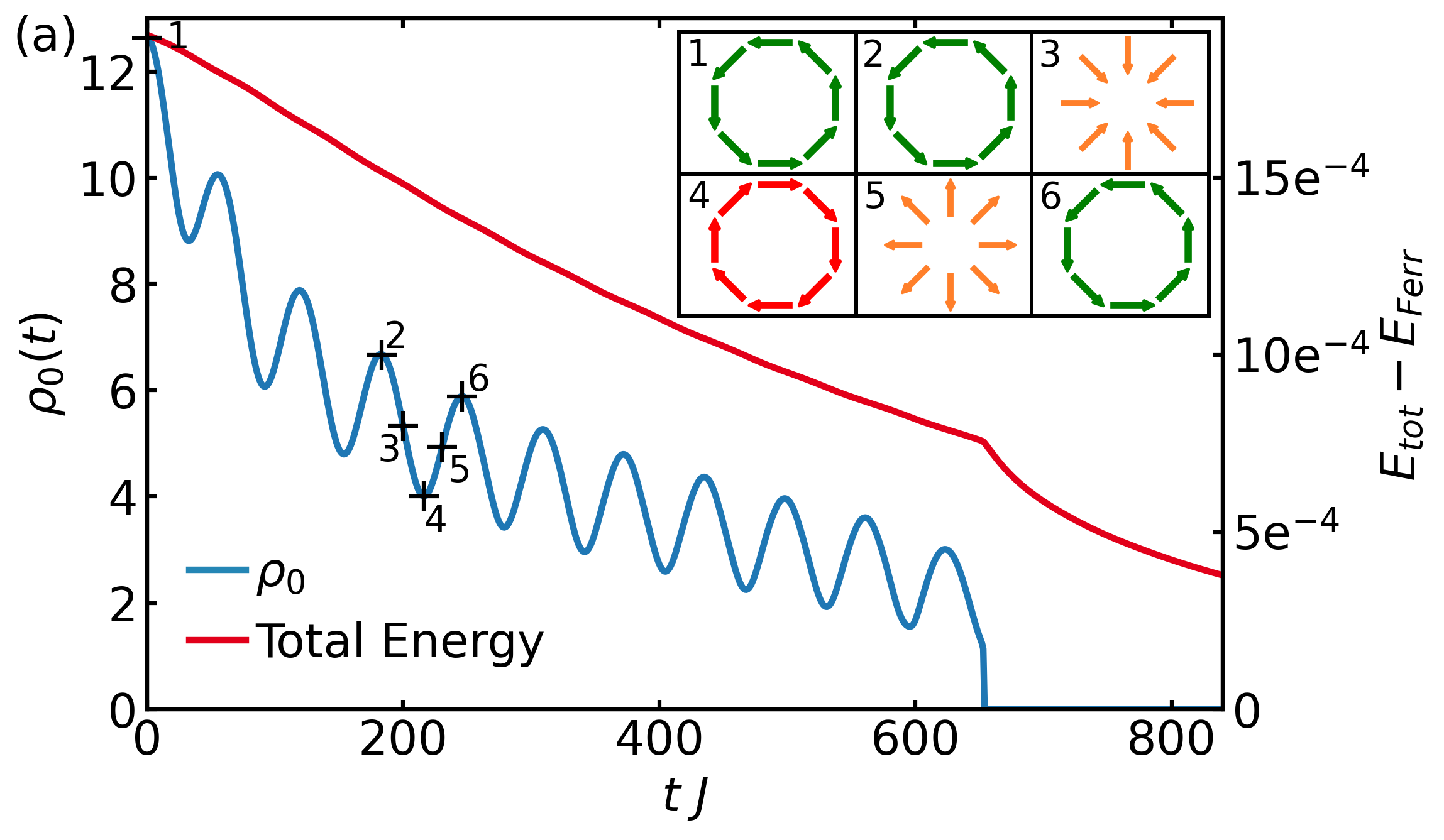}
	\includegraphics[width=0.47\textwidth]{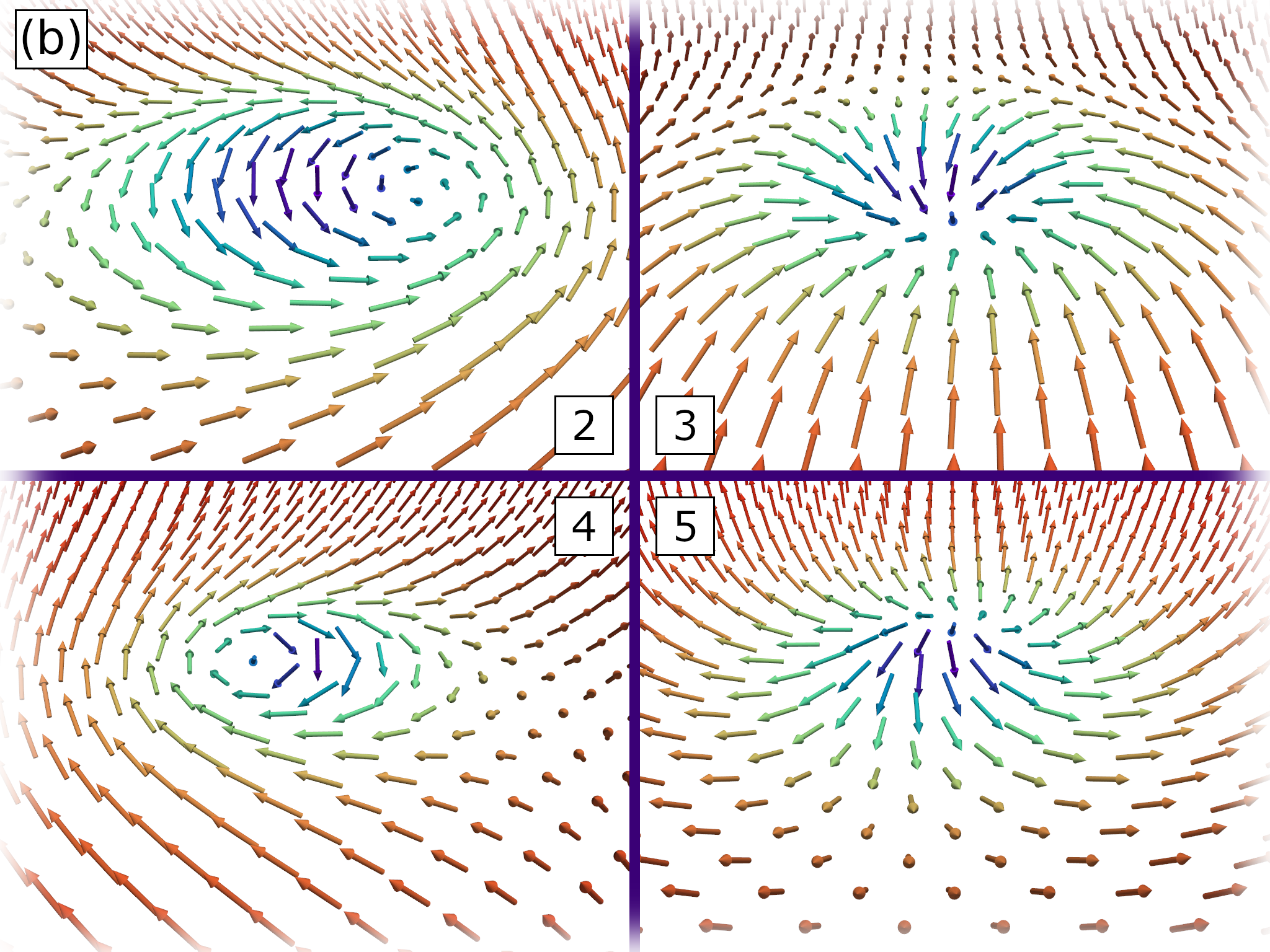}
	\caption{(a) The time-dependent skyrmion radius (blue line) is shown together with the time evolution of the total energy (red line) for $D=0.08$, $B=0.1$, $\alpha=0.01$ and $a=0.8$. Snapshots of the skyrmion configuration at different points of time are shown in panel (b), together with the corresponding configurations in the upper right inset of panel (a). The configuration in green color is the energetically most favorable configuration of a Bloch skyrmion for the chosen parameter set, while the orange ones are the anti-N\'eel (3) and N\'eel (5) skyrmions that are higher in energy. The skyrmion with the highest energy is the anti-Bloch (4) skyrmion depicted in red.
		\label{fig:interneModen}}
\end{figure}

To further illustrate the internal transitions between the different skyrmion types due to the rotational mode, the temporal behavior of the spin configurations are shown in Fig.\ \ref{fig:interneModen}. Different skyrmion configurations are characterized by the skyrmion helicity $\varphi_0$. At $t=0$ (1), the skyrmion is in its energetically favorable Bloch-type configuration, $\varphi_0=\pi/2$. With time evolution the skyrmion helicity $\varphi_0(t)$ increases and reaches again $\pi/2$-values plus multiples of $2\pi$, causing the radius to reach periodically again a local maximum (2) where the skyrmion is in the Bloch-type configuration. However, with slightly increasing $\varphi_0(t)$ the radius $\rho_0(t)$ decreases, consistent with Eq.\ \eqref{eq:TriangularEnergy}, approaching the midpoint between a local maximum and a local minimum (3). It thereby undergoes a transition to the energetically less favorable anti-N\'eel-type configuration, $\varphi_0=\pi$. In the local minimum of the radius (4), finally, the skyrmion acquires an anti-Bloch-type configuration, $\varphi_0=3\pi/2$. Following Eq.\ \eqref{eq:TriangularEnergy}, this is the configuration with the highest energy w.r.t.\ $\varphi_0$. This precession-like increase of $\varphi_0(t)$, of angular velocity proportional to the strength of the external magnetic field, leading to the oscillations of $\rho_0(t)$, then renews, (5) and (6). On longer time scales, the energy decreases exponentially due to the Gilbert damping, causing on average an overall exponential decrease of $\rho_0(t)$. All these findings are in agreement with the coupled differential equations from our analytic model.

\begin{figure}[t!]
	\centering
	\includegraphics[width=0.47\textwidth]{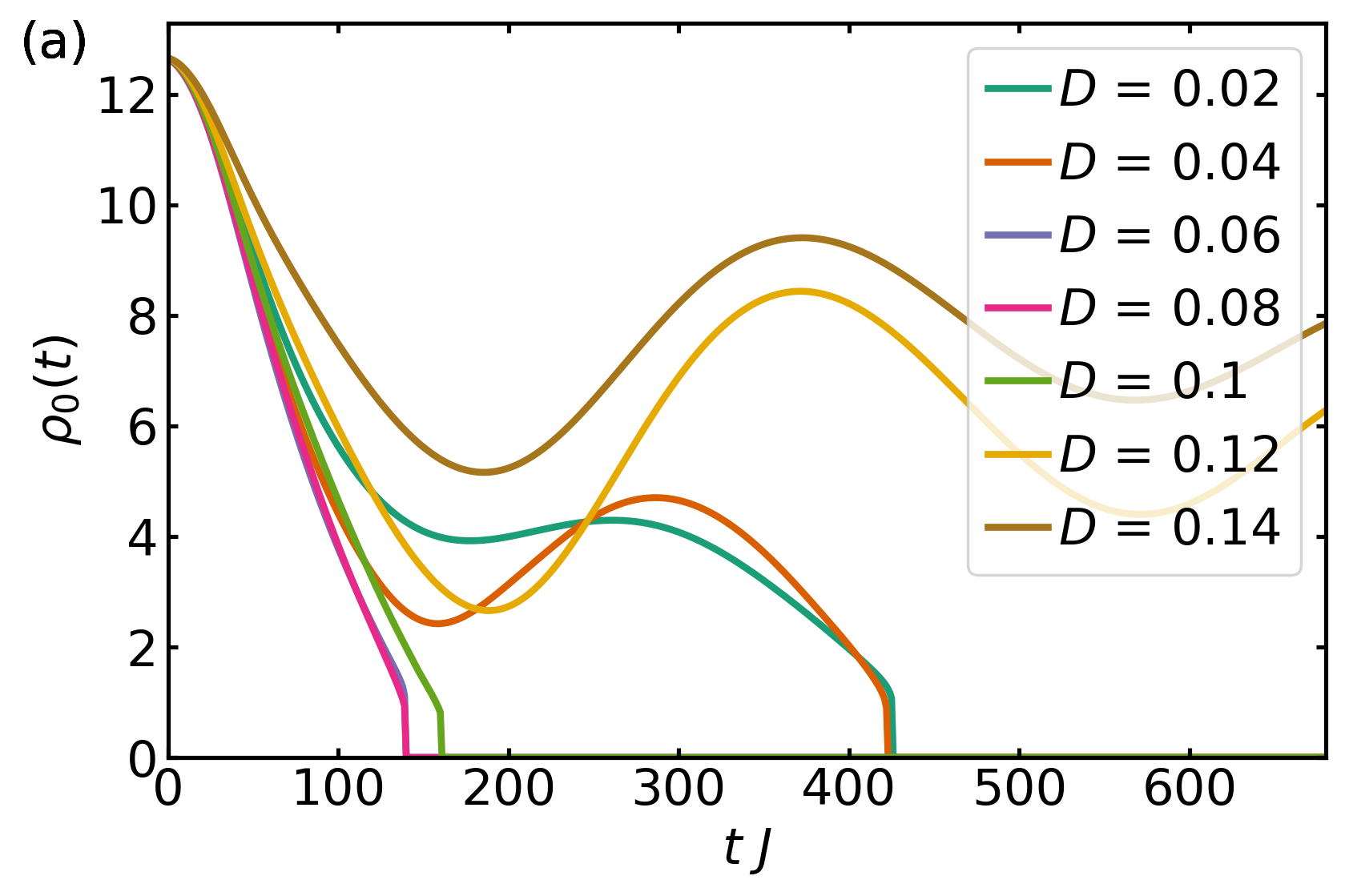}
	\includegraphics[width=0.47\textwidth]{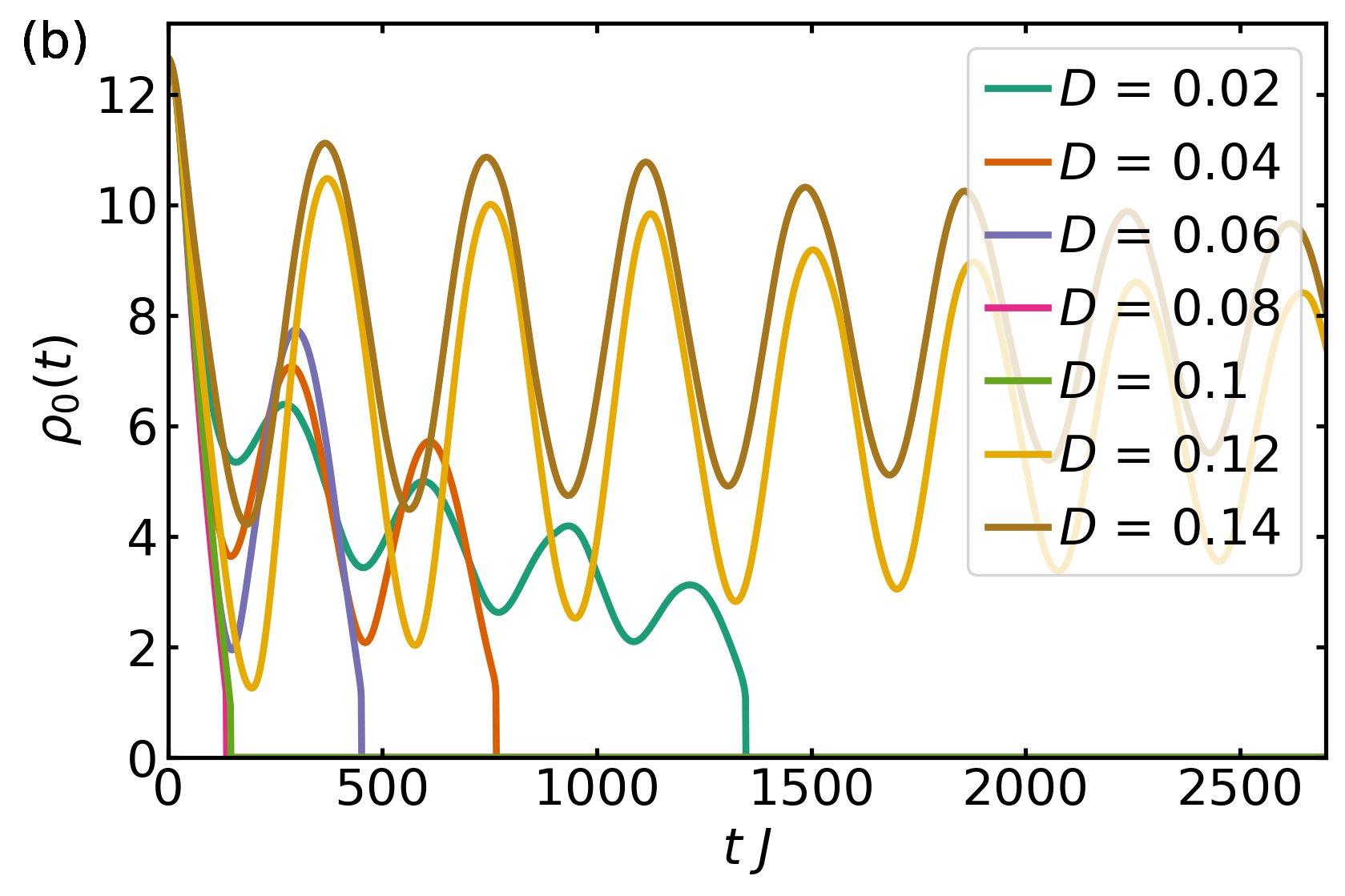}
	\caption{Time evolution of the skyrmion radius for varying values of the DMI and for a Gilbert damping of $\alpha=0.1$ (a) and $\alpha=0.01$ (b) for  $B=0.02$ for a lattice constant of $a=0.8$. The larger the DMI is, the faster the skyrmion shrinks until a DMI is reached that stabilizes the skyrmion at a finite radius.
		\label{fig:rho0_num_B0.02_tuneD}}
\end{figure}

%\subsection{Decay at varying values of the DMI}
Finally, we investigate the dependence of the skyrmion decay on the DMI. The time evolution of the skyrmion radius for varying DMI strengths is shown in Fig.\ \ref{fig:rho0_num_B0.02_tuneD} for an external magnetic field of $B=0.02$, a damping of (a) $\alpha=0.1$ and (b) $\alpha=0.01$ with a lattice constant $a=0.8$. With increasing DMI, the skyrmion decays faster until it reaches a value that stabilizes the skyrmion ($D \geq 0.12$) at a finite asymptotic radius $\rho_0(\infty)$. The larger the DMI is, without being strong enough to stabilize the skyrmion, the larger the amplitude of the rotational mode. In Eq.\ \eqref{eq:finalODErho} this is a consequence of the sinusoidal terms proportional to the DMI, acting as the amplitude. Thus, the radius is more probable to reach the value of the lattice constant at an earlier oscillation period as compared to a smaller DMI, which will trigger the skyrmion collapse earlier as well. The theoretical model does not include a lattice. Therefore, this collapse condition does not occur. With the lattice, increasing DMI will result in an even earlier collapse of the skyrmion radius. When the DMI does not stabilize the skyrmion, the same transitions through the different skyrmion configurations can be identified as examined in Fig.\ \ref{fig:interneModen}. When the DMI is large enough to stabilize the skyrmion at a finite radius, however, the skyrmion will stay in the Bloch-type configuration, accompanied by internal breathing modes. In this parameter regime, the amplitude of the breathing mode decreases with increasing DMI strength because the energetically optimal radius gets closer to the size of the initial configuration. Consequently, the skyrmion radius performs damped breathing oscillations around this minimal energy configuration until the final configuration is reached.

\begin{figure}
	\centering
	\includegraphics[width=0.47\textwidth]{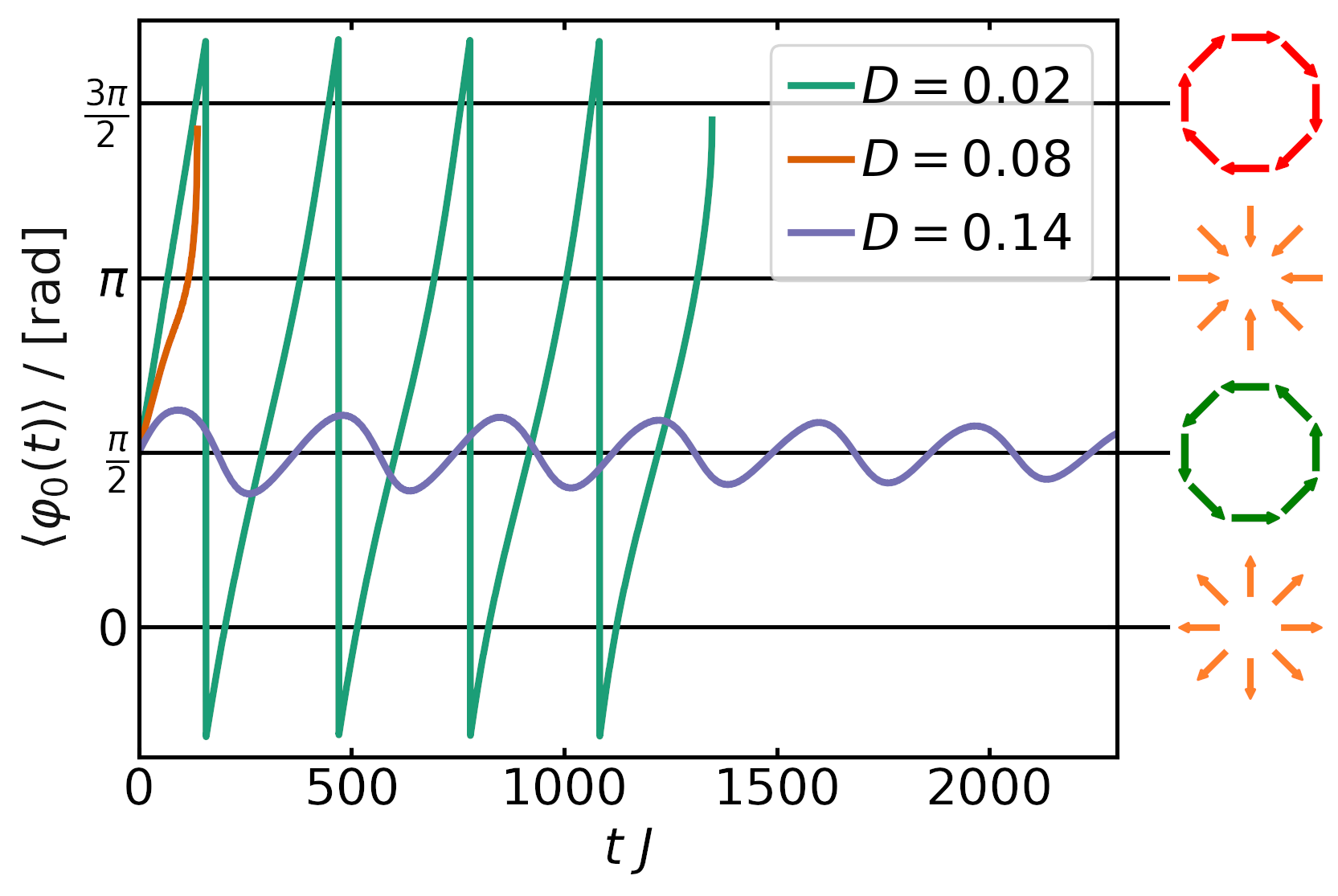}
	\caption{Time evolution of the averaged skyrmion helicity $\langle\varphi_0(t)\rangle$, obtained from the simulations depicted in Fig.\ \ref{fig:rho0_num_B0.02_tuneD} (b) for $D=0.02$, $0.08$ and $0.14$. When the DMI is too small to stabilize the skyrmion ($D=0.02$, $D=0.08$), the initial $\varphi_0$-value decays with time until the skyrmion collapses and the excess energy dissipates through magnon waves where the skyrmion's collective phase is lost. Before the skyrmion collapses, it rotates and periodically switches between the different skyrmion configurations, illustrated to the right of the graph. When the DMI is sufficiently strong to stabilize the skyrmion ($D=0.14$), it maintains a Bloch-type configuration throughout its breathing motion until it reaches its energetic minimum.
		\label{fig:PhiAngle}}
\end{figure}

The oscillatory behavior of the skyrmion helicity in the different parameter regimes also shows up in Fig.\ \ref{fig:PhiAngle}. The plot depicts the time evolution of the averaged helicity $\langle\varphi_0(t)\rangle$ for various DMI strengths. Respective configurations of the skyrmion are indicated by the black horizontal lines, with corresponding illustrations provided to the right of the graph. The averaging is performed at each time step over all $\varphi_{0i}$ angles on the grid as
\begin{equation}
\langle \varphi_0 \rangle = \frac{\sum_{i} \mathcal{W}_i \cdot \varphi_{0i}}{\sum_{i}\mathcal{W}_i},
\end{equation}
where $\varphi_{0}$ is weighted using the function $\mathcal{W}_i=1-\left|n_i^z\right|$. This way, magnetic moments pointing almost parallel or anti-parallel to the $z$-axes contribute less to the mean since their azimuthal angle is less sharp defined. After collapse, when the energy dissipates through magnon waves, the skyrmion's collective phase is lost.

	\section{Summary}
We derive and analyse a continuum model to describe the dynamics of skyrmion shrinking up to the eventual collapse. By employing a triangular-shaped skyrmion profile for the polar angle, we formulated a set of coupled nonlinear differential equations for the skyrmion radius and the helicity. At zero DMI the equation for the radius decouples from the one for the helicity, allowing for an analytic solution for $\rho_0(t)$ at $\tilde{D}=0$. Complementing our theoretical model, numerical simulations on a square lattice have been carried out for comparison.

Contrary to the commonly expected simple exponential decrease of the skyrmion radius, we reveal a more complicated time dependence. The radius crosses over from an exponential decay towards a square root decrease, $\sim (t - t_c)^{1/2}$, at small radii. We identify a critical time $t_c$ at which the skyrmion radius sharply drops to zero, depending logarithmically on the lattice constant as well as on the strength of the external magnetic field. Numerical solution of the coupled differential equations reveals, as supported by simulations, that finite values of the DMI shorten this critical time for the collapse. This is related to a continuous precession of the skyrmion phase taking place through its helicity $\varphi_0(t)$, which leads to cyclic transformations between Bloch-, anti-N\'eel-, anti-Bloch-, and N\'eel-type skyrmions. Due to energy conservation this rotational mode at finite DMI enforces oscillations of the skyrmion radius which then superimpose the exponential into square-root decay. Frequency and amplitude of this rotational mode are proportional to the external magnetic field strength and the DMI strength, respectively.

At sufficiently large DMI, the rotation through the different skyrmion phases stops and, for our choice of the DMI, a Bloch-type skyrmion of finite size is stabilized at $t\rightarrow\infty$. In this case, we find damped harmonic breathing oscillations around the minimum energy configuration. The frequency of these breathing modes clearly differ from the frequency of the rotational mode.

	\begin{acknowledgments}
	We would like to thank Martin Stier for fruitful discussions on the stability of ferromagnetic skyrmions and also Volodymyr Kravchuk and Markus Garst for inspiring discussions on the helicity of skyrmions. This work has been funded by the Deutsche For\-schungs\-gemeinschaft - Projektnummer 403505707 within the DFG Priority Program SPP 2137 ``Skyrmionics'' (Projektnummer 360506545).
	\end{acknowledgments}
	
%	\appendixSPP 2137:  Skyrmionics: Topologische Spin-Phänomene im Realraum für Anwendungen
%	\section{Section}
%	\label{apendix1}


\begin{thebibliography}{2}
\bibitem{TSkyrme} T. H. R. Skyrme, A non-linear field theory, Proc. R. Soc. Lond. A \textbf{260}, 127 (1961).

\bibitem{Moriya} T. Moriya, Anisotropic superexchange interaction and weak ferromagnetism, Phys. Rev. \textbf{120}, 91 (1960).

\bibitem{Tomasello} R. Tomasello, E. Martinez, R. Zivieri, L. Torres, M- Carpentieri, and G. Finocchio, A strategy for the design of Skyrmion racetrack memories, Sci. Rep. \textbf{4}, 6784 (2014).

\bibitem{Nagaosa} N. Nagaosa and Y. Tokura, Topological properties and dynamics of magnetic skyrmions, Nat. Nanotechnol. \textbf{8}, 899 (2013).

\bibitem{Leonov} A. O. Leonov, T. L. Monchesky, N. Romming, A. Kubetzka, A. N. Bogdanov, and R. Wiesendanger, The properties of isolated chiral skyrmions in thin magnetic films, New J. Phys. \textbf{18}, 065003 (2016).

\bibitem{Zhang} X. Zhang, M. Ezawa, and Y. Zhou, Magnetic skyrmion logic gates: conversion, duplication and merging of skyrmions, Sci. Rep. \textbf{5}, 9400 (2015).

\bibitem{Finocchio} G. Finocchio, F. Büttner, R. Tomasello, M. Carpentieri, and M. Kläui, Magnetic skyrmions: from fundamental to applications, J. Phys. D Appl. Phys. \textbf{49}, 423001 (2016).

\bibitem{Jiang} W. Jiang, G. Chen, K. Liu, J. Zang, S. G. E. te Velthuis, A. Hoffmann, Skyrmions in magnetic multilayers, Phys. Rep. \textbf{704}, 1 (2017).

\bibitem{RommingWiesendangerWritingDeleting} N. Romming, C. Hanneken, M. Menzel, J. E. Bickel, B. Wolter, K. von Bergmann, A. Kubetzka, and R. Wiesendanger, Writing and deleting single magnetic skyrmions, Science \textbf{341}, 636 (2013).

\bibitem{Stier} M. Stier, W. Häusler, T. Posske, G. Gurski, and M. Thorwart, Skyrmion-Anti-Skyrmion Pair Creation by in-Plane Currents, Phys. Rev. Lett. \textbf{118}, 267203 (2017).

\bibitem{Iwasaki} J. Iwasaki, M. Mochizuki, and N. Nagaosa, Universal current-velocity relation of Skyrmion motion in chiral magnets, Nature Commun. \textbf{4}, 1463 (2013).

\bibitem{BesserabGNEB} P. F. Bessarab, V. M. Uzdin, and H. Jónsson, Method for finding mechanism and activation energy of magnetic transitions, applied to skyrmion and antivortex annihilation, Comput. Phys. Commun. \textbf{196}, 335 (2015).

\bibitem{Rybakov} F. N. Rybakov, A. B. Borisov, S. Blügel, and N. S. Kiselev, New Type of Stable Particlelike States in Chiral Magnets, Phys. Rev. Lett. \textbf{115}, 117201 (2015).

\bibitem{Lobanov} I. S. Lobanov, H. Jónsson, and V. M. Uzdin, Mechanism and activation energy of magnetic skyrmion annihilation obtained from minimum energy path calculations, Phys. Rev. B \textbf{94}, 174418 (2016).

\bibitem{HeilRoschMasell} B. Heil, A. Rosch, and J. Masell, Universality of annihilation barriers of large magnetic skyrmions in chiral and frustrated magnets, Phys. Rev. B \textbf{100}, 134424 (2019).

\bibitem{Vlasov} S. M. Vlasov, P. F. Bessarab, I. S. Lobanov, M. N. Potkina, V. M. Uzdin, and H. Jónsson, Magnetic skyrmion annihilation by quantum mechanical tunneling, New J. Phys. \textbf{22}, 083013 (2020).

\bibitem{Derras-ChoukChudnovskyGaraninQuantum} A. Derras-Chouk, E. M. Chudnovsky, and D. A. Garanin, Quantum collapse of a magnetic skyrmion, Phys. Rev. B \textbf{98}, 024423 (2018).

\bibitem{Braun} H.-B. Braun, Fluctuations and instabilities of ferromagnetic domain-wall pairs in an external magnetic field, Phys. Rev. B \textbf{50}, 16485 (1994).

\bibitem{KubetzkaWiesendanger} A. Kubetzka, O. Pietzsch, M. Bode, and R. Wiesendanger, Spin-polarized scanning tunneling microscopy study of 360° walls in an external magnetic field, Phys. Rev. B \textbf{67}, 020401 (2003).

\bibitem{RommingWisendangerDomainWall} N. Romming, A. Kubetzka, C. Hanneken,
K. von Bergmann, and R. Wiesendanger, Field-Dependent Size and Shape of Single Magnetic Skyrmions, Phys. Rev. Lett. \textbf{114}, 177203 (2015).

\bibitem{Wang} X. S. Wang, H. Y. Yuan, and X. R. Wang, A theory on skyrmion size, Commun. Phys. \textbf{1}, 31 (2018).

\bibitem{RohartMiltatThiaville} S. Rohart, J. Miltat, and A. Thiaville, Path to collapse for an isolated N\'eel skyrmion, Phys. Rev. B \textbf{93}, 214412 (2016).

\bibitem{CaiChudnovskyGaranin} L. Cai, E. M. Chudnovsky, and D. A. Garanin, Collapse of skyrmions in two-dimensional ferromagnets and antiferromagnets, Phys. Rev. B \textbf{86}, 024429 (2012).

\bibitem{Derras-ChoukChudnovskyGaranin} A. Derras-Chouk, E. M. Chudnovsky, and D. A. Garanin, Dynamics of the collapse of a ferromagnetic skyrmion in a centrosymmetric lattice, Phys. Rev. B \textbf{105}, 134432 (2022).

\bibitem{MuehlbauerBoeni} S. Mühlbauer, B. Binz, F. Jonietz, C. Pfleiderer, A. Rosch, A. Neubauer, R. Georgii, and P. Böni, Skyrmion Lattice in a Chiral Magnet, Science \textbf{323}, 915 (2009).

\bibitem{LandauLifshits} L. Landau, and E. Lifshits, On the theory of the dispersion of magnetic permeability in ferromagnetic bodies, Phys. Z. Sowjet. \textbf{8}, 153 (1935).

\bibitem{Gilbert} T. L. Gilbert, A Phenomenological Theory of Damping in Ferromagnetic Materials, IEEE Trans. Magn. \textbf{40}, 3443 (2004). 

\bibitem{BogdanovYablonskii} A. N. Bogdanov, and D. A. Yablonskii, Thermodynamically stable ``vortices'' in magnetically ordered crystals. The mixed state of magnets, Sov. Phys. JETP, \textbf{68}, 101 (1989).

\bibitem{McKeeverEverschor-Sitte} B. F. McKeever, D. R. Rodrigues, D. Pinna, Ar. Abanov, Jairo Sinova, and K. Everschor-Sitte, Characterizing breathing dynamics of magnetic skyrmions and antiskyrmions within the Hamiltonian formalism, Phys. Rev. B \textbf{99}, 054430 (2019).

\bibitem{KronmuellerFaehnle} H. Kronmüller, and M. Fähnle, {\em Micromagnetism and the Microstructure of Ferromagnetic Solids}, (Cambridge University Press, Cambridge, 2003).

\bibitem{Schuette} C. Schütte, and M. Garst, Magnon-skyrmion scattering in chiral magnets, Phys. Rev. B \textbf{90}, 094423 (2014).

\bibitem{Everschor-Sitte} K. Everschor-Sitte, J. Masell, R. M. Reeve, and M. Kläui, Perspective: Magnetic skyrmions --- Overview of recent progress in an active research field, J. Appl. Phys. \textbf{124}, 240901 (2018).

\bibitem{Albrecht24} M. Hassan, S. Koraltan, A. Ullrich, F. Bruckner, R. O. Serha, K. V. Levchenko, G. Varvaro, N. S. Kiselev, M. Heigl, C. Abert, D. Suess, and M. Albrecht, Dipolar skyrmions and antiskyrmions of arbitrary topological charge at room temperature, Nature Physics \textbf{20}, 615 (2024).

\bibitem{BogdanovHubert} A. Bogdanov, and A. Hubert, Thermodynamically stable magnetic vortex states in magnetic crystals, J. Magn. Magn. Mater. \textbf{138}, 255 (1994).

\bibitem{Gradshteyn} I. S. Gradshteyn, and I. M. Ryzhik, Table of Integrals, Series, and Products, 8th ed., D. Zwillinger, and V. Moll, (Academic Press, Amsterdam, 2015).

\bibitem{NavauSanchez} C. Navau, N. Del-Valle, and A. Sanchez, Analytical trajectories of skyrmions in confined geometries: Skyrmionic racetracks and nano-oscillators, Phys. Rev. B \textbf{94}, 184104 (2016).

\bibitem{SrivastavaBea} T. Srivastava, M. Schott, R. Juge, V. Křižáková, M. Belmeguenai, Y. Roussigné, A. Bernand-Mantel, L. Ranno, S. Pizzini, S.-M. Chérif \textit{et al}., Large-Voltage Tuning of Dzyaloshinskii–Moriya Interactions: A Route toward Dynamic Control of Skyrmion Chirality, Large-Voltage Tuning of Dzyaloshinskii–Moriya Interactions: A Route toward Dynamic Control of Skyrmion Chirality, Nano Lett. \textbf{18} 4871 (2018).

\bibitem{KoyamaChiba} T. Koyama, Y. Nakatani, J. Ieda, and D. Chiba, Electric field control of magnetic domain wall motion via modulation of the Dzyaloshinskii-Moriya interaction, Sci. Adv. \textbf{4}, eaav0265 (2018).

\bibitem{KatoHayashi} N. Kato, M. Kawaguchi, Y.-C. Lau, T. Kikuchi, Y. Nakatani, and M. Hayashi, Current-Induced Modulation of the Interfacial Dzyaloshinskii-Moriya Interaction, Phys. Rev. Lett. \textbf{122}, 257205 (2019).

\bibitem{BelavinPolyakov} A. A. Belavin, and A. M. Polyakov, Metastable states of two-dimensional isotropic ferromagnets, Pis'ma Zh. Eksp. Teor. Fiz. \textbf{22}, 503 (1975).

\end{thebibliography}
\end{document}